\documentclass[aps,prb,showpacs,floatfix,twocolumn,superscriptaddress]{revtex4-2}
\usepackage{amsmath} 
\usepackage{bbding}
\usepackage{pifont}
\usepackage{sublabel}
\usepackage{amssymb}
\usepackage{mathrsfs}
\usepackage{amsmath}
\usepackage{amssymb,stackengine}
\usepackage{pifont}
\usepackage{enumitem}
\usepackage{epstopdf}
\usepackage[ 
colorlinks=true,
pdfborder={0 0 0},
linkcolor=red,
citecolor = red
]{hyperref}
\usepackage{graphicx}
\usepackage{physics}
\usepackage{bm,amsmath,amssymb,accents}

\usepackage{dcolumn}
\usepackage{xcolor}
\usepackage{bm}

\newcommand{\tcr}{\textcolor{black}}

\newcommand{\tony}[1]{\textcolor{red}{[ #1]}}
\usepackage{algorithm}
\usepackage{algorithmicx}
\usepackage{algpseudocode}
\usepackage{placeins} 
\makeatletter

\newcommand{\Rmnum}[1]{\expandafter\@slowromancap\romannumeral #1@}
\makeatletter
\newcommand{\circdot}{%
  \mathrel{%
    \vphantom{\circlearrowleft}%
    \mathpalette\circd@t\relax
  }%
}
\newcommand{\circd@t}[1]{{\ooalign{%
            $\m@th#1\circlearrowleft$\cr
            \hidewidth$\m@th#1\cdot$\hidewidth\cr}}}
\makeatother

\begin{document}


\title{First-Order Vortex Lattice Melting in Bilayer Ice: A  Monte Carlo Method Study}
\author{Telun Zhong}
\affiliation{College of Physics, Taiyuan University of Technology,
    Shanxi 030024, China}

\author{Heyang Ma}
\affiliation{College of Physics, Taiyuan University of Technology,
    Shanxi 030024, China}

\author{Peijun Zheng}
\affiliation{ Key Laboratory of Materials Physics, Institute of Solid State Physics, HFIPS, Chinese Academy of Sciences, Hefei 230031, China }
\author{Jie Zhang}
\thanks{zhangjie01@tyut.edu.cn}
\affiliation{College of Physics, Taiyuan University of Technology,
    Shanxi 030024, China}

\author{Wanzhou Zhang}
\thanks{zhangwanzhou@tyut.edu.cn}
\affiliation{College of Physics, Taiyuan University of Technology,
    Shanxi 030024, China}

\affiliation{
    Hefei National Laboratory for Physical Sciences at the Microscale and Department of Modern Physics, University of Science and Technology of China, Hefei 230026, China}
\date{\today }
\begin{abstract}
    Inspired by the stable bilayer water ice grown in the laboratory (Nature 577, 60, 2020), we propose a model representing water ice as a two-layer six-vertex model. Using the loop update Monte Carlo method, we unveil meaningful  findings. While the square lattice six-vertex model exhibits an antiferromagnetic to 
disordered phase transition known as the Berezinskii-Kosterlitz-Thouless  transition, we observe a different scenario for the bilayer six-vertex model, where the transition type transforms into an Ising transition.
    We discover the emergence of vortices in the disordered phase, and to stabilize them,  vortex excitation is induced. This leads to the presence of distinct 1/2 filling and 2/3 filling vortex lattice phases. \tcr{More importantly}, we identify the phase transitions between the vortex lattice phase and the \tcr{disordered} phase, as well as between the 1/2 and 2/3 vortex lattices, as being of first order.
    \tcr{We also propose an experimental scheme for realizing a two-layer six-vertex model based on the artificial ice of particles in a double well trap array.}
    Our findings provide valuable \tcr{insight} into the nature of phase transitions occurring in layered water ice and artificial spin ice systems in experimental setups.

\end{abstract}
\maketitle


\section{introduction}





Ice is a common substance in nature.
There are various types of ice, including the solid form of liquid  water~\cite{Salzmann2021}, spin ice in real materials~\cite{Steven}, artificial spin ice~\cite{arm, RevModPhys.85.1473}, and particle ice~\cite{cj7}.
One common feature of the different forms of ice is the ice rule,  i.e., the so called
two-in (close) two-out (far away) topological constraint, introduced by  Pauling in 1935~\cite{LinusCarlPauling}.
Water ice exhibits 19 stable geometric structures, currently identified through  high pressure and low temperature experiments~\cite{Salzmann2021}.
Spin ice also exists \tcr{many} real materials with different structures, such as rare-earth pyrochlores~$\text{Ho}_2\text{Ti}_2\text{O}_7$~\cite{Harris}.

Researchers have also attempted to grow artificial spin ices~\cite{PhysRevLett.129.057202}, due to their controllability.
The microscale systems used to create artificial spin ice typically involve magnetically interacting nanoislands
or nanowire links\cite{Wang2006}, superconducting-qubit arrays~\cite{AndrewD}.
The physics studied through ice \tcr{is very broad}; \tcr{it covers} residual entropy ~\cite{Ramirez}, frustration \cite{Steven}, monopoles~\cite{Castelnovo} and so \tcr{forth}.


Artificial spin ice can construct various  configurations, including \tcr{the} vortex lattice (VL) phase~\cite{PhysRevLett.111.067001}. \tcr{Its} periodic structure composed of vortices is called the VL. As vortex states can help understand superconductors~\cite{PhysRevB.96.134515},
\tcr{many studies} have been done on  transitions between  vortex lattices and other phases~\cite{firstordervortex, firstordervortex2, Zeldov1995ThermodynamicOO,PhysRevLett.80.4297}.
In the XY model, the unbinding of \tcr{the} vortex-antivortex pairs is considered the cause of \tcr{the} Berezinskii-Kosterlitz-Thouless (BKT) phase transition~\cite{xy1,xy2,xy33}. However,  the vortex lattice leads to a first order phase transition in \tcr{the} real crystals
YBCO and $\text{YBa}_2\text{Cu}_3\text{O}_7$
~\cite{firstordervortex, firstordervortex2, Zeldov1995ThermodynamicOO,PhysRevLett.80.4297}.

\begin{figure}[t]
    \includegraphics[width=0.35\textwidth]{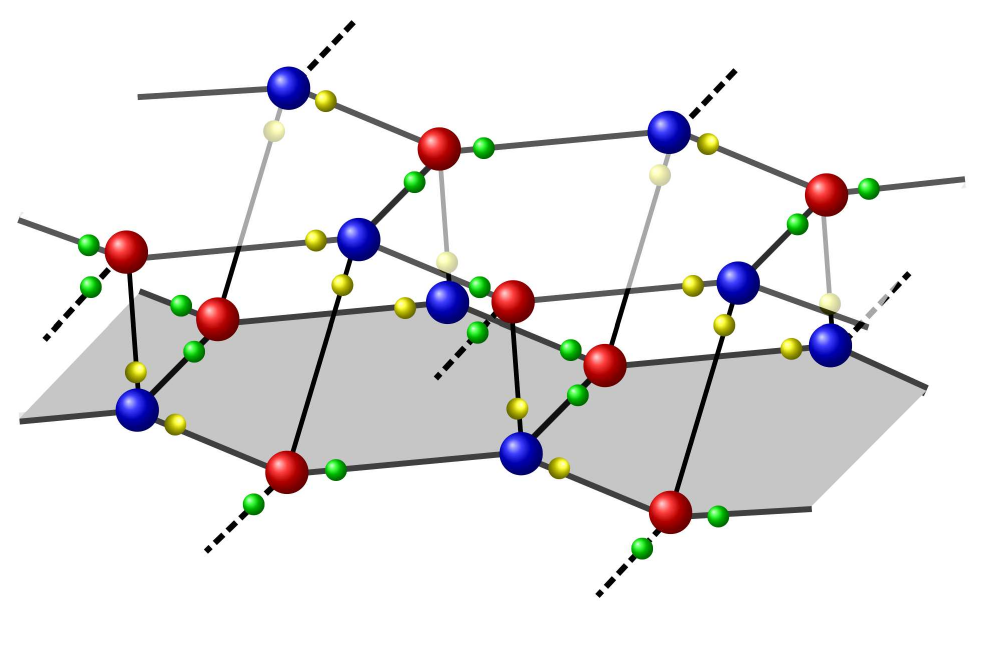}
    \vskip -0.3cm
    \caption{
        The construction of the two-layer ice model.
        The large balls (in red and blue) represent
        $O^{2-}$s and the small balls denote
        $H^+$s. The different colors for the large ball means AB stacking~\cite{AAice}. The ``two-in-two-out rule can be demonstrated by the small balls in different colors.}
    \label{fig:latt2}
\end{figure}

Of course, the ice physics can also be explored through experiments with water ice. In 2020, the group  at Peking University \tcr{confirmed} the existence of two-dimensional two-layered water ice~\cite{jiangying}.  The positions of the upper layer oxygen ions, as well as the connections between the oxygen ions, are exactly the same as those in the lower layer, i.e., AA stacking ice. Such \tcr{a} stable structure of water was first predicted in 1997 using \tcr{a} molecular dynamics simulation~\cite{Koga}.
Actually, there exists another type of ice with an AB stacking structure~\cite{AAice} as shown  Fig.~\ref{fig:latt2}. %
There is a relative 180 degree rotation between the two layers, which are connected by hydrogen bonds. Alternatively, the hydrogen bonds of the first layer are shifted onto the faces of the second layer. It has been confirmed to be stable under reasonable \tcr{temperatures and pressures} by \tcr{\em ab-initio} calculations~\cite{AAice} \tcr{though lacking in} experimental preparation~\cite{jiangying}.

To inspire experimental physicists to achieve AB or other \tcr{types} of bilayer ice in the future, we
further convert  \tcr{the} bilayer ice to \tcr{the} bilayer six-vertex (6V) \tcr{model}~\cite{22in}.
On the basis of the bilayer square lattice 6V model,  the  AB stacking  honeycomb 6V model can be formed  by  bending one leg of the \tcr{vertices} to the appropriate degree. 
Hydrogen ions and oxygen ions in ice can be close \tcr{to each other}  or far apart, similar to 
\tcr{the situation of} spin-up and spin-down.
Viewing ice as a spin system helps understand the phase transitions between different types of ice under various conditions like temperature and pressure.
    {\it It is not clear \tcr{whether or not there is} something new in the bilayer honeycomb 6V  model.}




In this paper,  we apply a large-scale loop Monte Carlo (MC) simulation  to study the proposed  6V model. By properly defining and scanning  the type and weight, including the vortex exciting weight, we explore the phase \tcr{diagrams} systematically.  The system includes ferromagnetic, antiferromagnetic, 1/2-filling VL, 2/3-filling VL, and disordered phases.
The theoretically discovered types of phase transitions, such as first order phase transitions, also \tcr{provide insight} into understanding previous experiments~\cite{firstordervortex, firstordervortex2, Zeldov1995ThermodynamicOO,PhysRevLett.80.4297}.


The outline of this work is as follows. Sec.~\ref{sec:model} introduces the bilayer 6V model, algorithm, and the measured quantities.  Sec.~\ref{sec:res1}  describes the phase diagram and details without the effects of vortex weight $V=1$.
Sec.~\ref{sec:vortexne0}  describes the phase diagram with the effects of vortex weight $V \ne 1$. \tcr{Sec.~\ref{sec:ex} discusses the experimental realization.}
Conclusive comments and outlook are made in Sec. \ref{sec:conclusion}.

Physically, apart from the bilayer 6V model initially \tcr{proposed}, we have made new discoveries as follows:

(I) The transition between the antiferromagnetic phase to the disordered phase is \tcr{of} Ising type for our bilayer 6V model.
However, for the 6V  model on the square lattice,  the transition \tcr{is of} BKT type~\cite{Fmodel}.

(II) Two types of vortex lattice phase \tcr{are} found when vortex excitation is induced. The transition from  vortex lattice phases  to other phases is  of first order, consistent with previous crystal experiments~\cite{firstordervortex,firstordervortex2}
. \tcr{This} helps researchers understand that not all phase transitions involving vortices are BKT phase transitions.

On the algorithmic level, although Ref.~\cite{22in} has simulated the single-layer planar 6V model using \tcr{the} loop algorithm, \tcr{our model} is the non-planar 6V model and we provide the details of closing loops.  In addition, we 
introduce a
\tcr{Metropolis type short-loop} update method for the purpose of ergodicity.

\tcr{Experimentally,
    bilayer water
    ice, where the ordering is due to the proton positions~\cite{jiangying}, is probably better captured by particle-based ice~\cite{cj7}.
    Here, we propose to groove or \tcr{imprint} both bilayer \tcr{lattices} onto a single-layer substrate, and each edge of the lattices are realized  by a double well trap with only one particle. }

\begin{figure}[h]
    \includegraphics[width=0.45\textwidth]{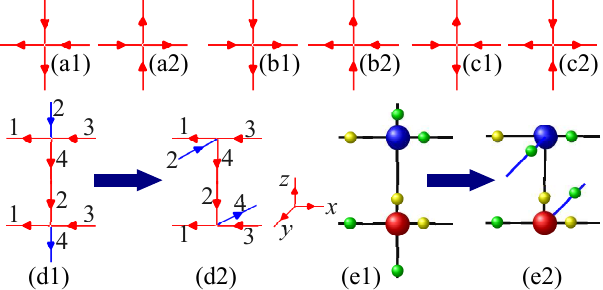}
    \includegraphics[width=0.45\textwidth]{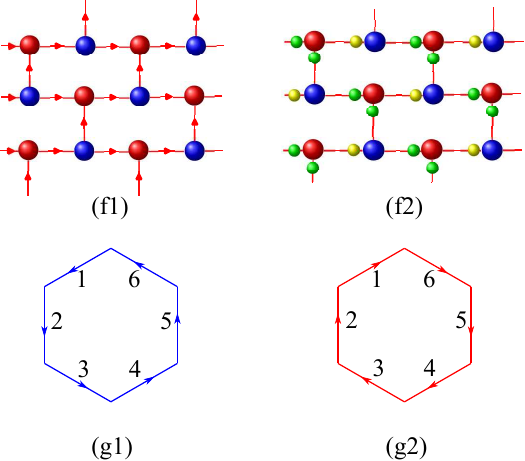}
    \caption{Schematic diagram of the 6V model and  water molecules. (a1)-(c2) The configurations of six vertices and their weights defined as $a$, $a$, $b$, $b$, $c$ and $c$, respectively.
        (d1)-(d2) Connecting the two vertices  by the two vertical red legs marked by 2 and 4.
        (e1)-(e2) Rotating the two legs,  \tcr{marked in blue}, on the horizontal plane.
        (f1)-(f2) Mapping to the structure of real water
        ice molecules. The small ball represents \tcr{the} hydrogen ion and the large ball represents \tcr{the} oxygen ion. (g1)-(g2) The definition of \tcr{the} vortex.
    }
    \label{fig:latt4}
\end{figure}
\section{Model, algorithm and quantities }
\label{sec:model}
\subsection{The 6V model with vortex weight}
\subsubsection{\tcr{Hamiltonian} and partition function}

Unlike the models such as the Ising\tcr{~\cite{lsing}}, XY~\cite{xy1,xy2,xy33}, Potts~\tcr{~\cite{Potts}} models or the coupled spins such \tcr{the} Ising-XY model~\cite{mhy}, etc., the famous 6V model does not have a explicit Hamiltonian.
\tcr{However,} each type of vertex has its own weight and can also have an equivalent energy, and the vertex satisfies the ``two-in-two-out'' topological constraints.
For convenience, to measure physical quantities related to specific heat and other energy-related quantities, a quasi-Hamiltonian is introduced as:


\begin{equation}
    \mathcal{H}  =-\sum_{i=1 }^{2N} \epsilon_{i} -v\sum_{ j=1 }^{2f} n_i,
    \label{eq:hami}
\end{equation}
where $\epsilon_{i}$ is the effective energy for each vertex labeled by $i$,  and $v$ is a local vortex energy for each plaquette.
$n_i$ represents the number of vortices in each plaquette,  where
$n_i=1$
denotes  a clockwise $\circlearrowright$ or a counterclockwise $\circlearrowleft$ vortex, and $n_i=0$ signifies the absence of a vortex.
$N$ is the total number of \tcr{vortices} of one layer and $f$ is \tcr{the} 
total \tcr{number of} faces of the honeycomb lattice of one layer.



Using the Boltzmann weight factor, the vertex weight is:
\begin{equation}
    \omega _i=\text{exp}(\beta\epsilon _i )
\end{equation}
and the vortex weight is:
\begin{equation}
    V _i=\text{exp}(\beta v n_i)
\end{equation}
where $\beta = 1/k_BT$ is the inverse temperature and set to 1, and the partition function of the system is defined as follows:

\begin{equation}
    \mathcal{Z}  =\prod_{i=1}^{2N} \omega_{i}\prod_{j=1}^{2f}V_j.
\end{equation}

Figures~\ref{fig:latt4} (a1)-(c2) show
the configurations of six \tcr{vertices}, and each vertex has four legs marked by the red arrows.
The directions of the legs satisfy two \tcr{of them} facing out away from the center and two legs pointing toward the center.
For simplicity, the weights of the six \tcr{vertices} take values as:

\begin{equation}
    w_i=[a, a, b, b, c, c] ~.
\end{equation}
\tcr{Here}, the flipping symmetry of the leg direction is considered, i.e., the  state of (a1) can be obtained by flipping the directions of the four legs of (a2), and therefore
the weights of these two \tcr{vertices} are the same as $a$.
\tcr{In total}, only three possible weight values $a$, $b$ and $c$ are needed for the six different configurations.
This tradition is also used in Ref.\tcr{~\cite{Fmodel}}.

It should be \tcr{noted} that the vortex is defined as shown in Figs.~\ref{fig:latt4} (g1) and (g2). 
Numerically, the requirement is that the angle difference between adjacent vectors is $\frac{\pi}{3}$, i.e.,
\begin{equation}
    \theta_{\text{mod \{i,6\}+1 }}-  \theta_{i} = \pi/3,
\end{equation}
where $i$ ranges from 1 to 6. This type of definition  does not need the saw function in  Ref.~\cite{mhy} and  is the same \tcr{as} the one in Ref.~\cite{ding}.


\subsubsection{The formation of two-layer honeycomb lattice structure}
Figures~\ref{fig:latt4}(d1)-(e2)
show how to use the
\tcr{vertices}  to construct \tcr{two-layer}  honeycomb lattices.
Firstly,
6V models are usually constructed in the two-dimensional plane, and it is rare to see two-layer 6V  models.
Here, the interlayer coupling
is realized by   connecting the two vertices,  using the two red legs marked 2, 4 in the vertical direction.
Then, the second step \tcr{rotates} the two legs \tcr{(in blue)} to the horizontal $y$-direction.
A small coordinate axes \tcr{system denoted} $x$, $y$, $z$ \tcr{is} shown for reference.



It is also essential to illustrate the relationship between the two-layer honeycomb 6V model and the structure of ordered water molecules. In Figs.~\ref{fig:latt4}(f1)-(f2), the arrows pointing towards the center of the vertex 
\tcr{corresponds to the} approaching of the hydrogen (H) ions \tcr{towards} the oxygen (O) ion in a real water molecule. Conversely, arrows pointing away from the center indicate that the hydrogen ions \tcr{stay} away  from the oxygen ion. Ultimately, the arrangement of vertices can describe the structure of real water ice molecules.



In addition, the configurations of the system also \tcr{depend on} the topological constraints of  \tcr{``two-in-two-out''} rules~\cite{22in, 22out}. In other words, the arrows in the vertices have two pointing \tcr{of them} to the center and the other two back to the center.
The real configurations of large lattices \tcr{need} to be simulated by various MC methods~\cite{Monte1,Monte2}, which \tcr{will be discussed in the} next section.

\begin{figure}[H]
    \includegraphics[width=0.45\textwidth, height= 0.2\textwidth]{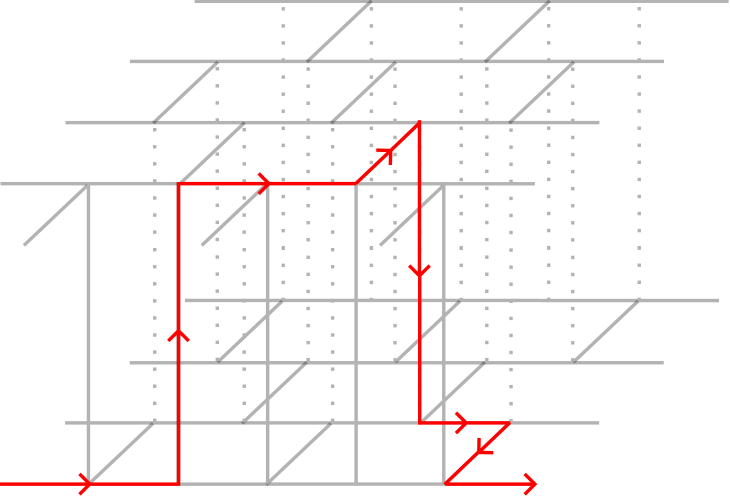}
    \caption{ Schematic diagram of the loop algorithm in a two-layer honeycomb lattice. The \tcr{starting} point is selected randomly, and the red arrows illustrate the path of the process. When the head of the loop \tcr{meets} the start vertex, the loop is closed and the configuration is updated.
        For convenience in defining the coordinates of the vertices, the honeycomb is defined as a topologically equivalent brick wall structure.}
    \label{fig:loop}
\end{figure}

\subsection{Methods and the measured quantities}
\label{sec:algorithm}
\subsubsection{Methods}

In this paper, we apply the loop algorithm, which \tcr{has} proven effective in studying various systems both classical systems  systems~\cite{22in},
and quantum systems~\cite{loop2,loop3}.
A similar loop algorithm is the famous worm algorithm~\cite{worm,worm2}, which involves a partial loop with two open ends with very efficient dynamical 
\tcr{behaviors}~\cite{d1}.
The 6V model is very similar to the flow representation of other models~\cite{flow1,flow2,flow3}.


To execute a loop update, the following steps are performed:

\begin{enumerate}
    \item
          First, we initialize the system with
          $N=2L_x\times L_y$  vertices, and then randomly select one of the vertices. Next, we randomly choose one of the four legs of the vertex to place the head of the loop.


    \item
          The leg where the head of the loop is located is used as the entrance leg, and then again one of the four legs from that vertex is chosen as the exit leg with a certain probability.


    \item
          The head of the loop enters the next new vertex, and the exit leg of the previous vertex is connected to this new vertex.


    \item
          To continue the process, we repeat steps 2 and 3 until the head of the loop and the end of the loop meet. Additionally, as the head of the loop traverses each leg, the state (arrows) of that leg should be flipped.


\end{enumerate}


Let's explain how to choose the exit leg when the ring head has been determined. Here, we employ the Metropolis-Hastings strategy. Suppose the weight of the reference vertex is $w_0$, and the weights of the new vertices resulting from exiting from the four legs are $w_1$, $w_2$, $w_3$, and $w_4$ respectively. We choose a random number between 0 and 1 to determine the interval in which the random number falls. These intervals are defined as $[0, p_1]$, $(p_1, p_1+p_2]$, $(p_1+p_2, p_1+p_2+p_3]$, and $(p_1+p_2+p_3, 1]$, where the probabilities are defined as:

\begin{equation}
    p_i= w_i/ \sum_{i=1}^4 w_i.
\end{equation}
If the outgoing and incoming legs are exactly the same, then there is no update, and the corresponding probability $p$ is referred to, as the bounce probability. Generally, a higher bounce probability leads to lower efficiency of the cluster algorithm~\cite{loop2,loop3}.

Therefore, analyzing the possible range of bounce \tcr{probabilities} under different parameters (see Table \ref{tab:my_label1}) \tcr{would} be helpful to ensure the feasibility of the code.

\begin{table}[h]
    \centering
    \begin{tabular}{c|c|c|c|c|c}
        \hline
        \hline
        ~~$a+b$~~    & $~\left | a-b \right |~ $ & ~~~$c$ ~~~ & ~~~~~$p_a$~~~~~      & ~~~~~$p_b$ ~~~~~     & ~$p_c$~                          \\
        \hline
        $<1$         & $\setminus $              & $1$        & $ (0,\frac{1}{2})$   & $ (0,\frac{1}{2})$   & $ (\frac{1}{2},1)$               \\
        \hline
        $\setminus $ & $>1$                      & $1$        & $ (0,\frac{a}{a+1})$ & $ (0,\frac{b}{b+1})$ & $ (\frac{1}{a+b+1},\frac{1}{2})$ \\
        \hline
        \hline
    \end{tabular}
    \caption{Parameters $a$, $b$, and $c$, and the possible range of bounce probability $p_a$, $p_b$ and $p_c$, where the subscripts $a$, $b$, $c$ represent the vertices of type $a$, $b$, $c$ that the loop encounters during its walk. }
    \label{tab:my_label1}
\end{table}



Loop close is an important step in the loop algorithm. In some cases, the length of the loop is very big and the code consumes  very long \tcr{run times}. \tcr{The authors of} Ref.~\cite{KaoYing-Jer} even cut the loop by the so-called short-loop \tcr{methods}.

\begin{figure}[h]
    \includegraphics[width=0.48\textwidth]{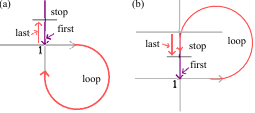}
    \caption{The two ways in which the loops close.  (a) \tcr{The first and last legs} overlap at the same leg, and (b) the last leg \tcr{connecting to} the first leg but the two \tcr{legs} belong to a pair of neighborhood \tcr{vertices}.}
    \label{fig:bihe}
\end{figure}

Here, we resort to the method dealing with the quantum Bose-Hubbard models~\cite{Lilaixi.134536,zwz1,Liran,yinruoxi}.  Two ways of closing loops are used.   In Fig.~\ref{fig:bihe} (a), the beginning leg \tcr{is} labeled by ``first" 
\tcr{and} the ending leg is marked by ``last". In this case, the first and the last legs meet (overlap) at the same leg, and then the loop \tcr{closes}.
In Fig.~\ref{fig:bihe} (b), the last leg connects to the position of the initial \tcr{leg}. This constraint arises from the ``2-in and 2-out" condition, allowing the loop closure only when connected to the initial leg.
The distinction lies in the fact that in the former case, the vertex 
labeled ``1'' undergoes two updates, whereas in the latter, it undergoes 
only \tcr{a} one time update.



\subsubsection{The measured quantities}
\noindent(\uppercase\expandafter{\romannumeral1}a)   The magnetization in the $x$ direction \tcr{is:}

\begin{equation}
    M^{\pm}_{x}=\sum_{i,j} (\pm 1)^{i+j}S_{i,j}^{x}/N,
\end{equation}
where  $(\pm 1)$ denotes the phase factor of ferromagnetization (FM) or antiferromagnetization (AFM) respectively. The \tcr{symbol $x$ refers} to the horizontal  directions of the lattice, and $i$ and $j$ \tcr{mean} the coordinates of the vertices in the $x$ and $y$ directions.



\noindent(\uppercase\expandafter{\romannumeral1}b)
The absolute values of magnetizations are:
\begin{equation}
    {m^{\pm}_{x}}={\langle |M^{\pm}_{x}|\rangle},
\end{equation}
where $\langle \rangle$ \tcr{signifies} the averages of Monte Carlo simulations.

\noindent(\uppercase\expandafter{\romannumeral1}c)
The striped \tcr{ferromagnetization} \tcr{is} defined as:
\begin{equation}
    m_s=\sum_{i=1}^{L}\left | \sum_{j=1}^{L} S_{i,j}^x \right |/N,
    \label{eq:ms}
\end{equation}
which is used later to define the striped specific heat $C_{ms}$.

\noindent(\uppercase\expandafter{\romannumeral1}d)
The Binder \tcr{ratios} are also defined as:
\begin{align}
    Q   & ={\langle {(M^{-}_{x})}^2\rangle}^2/{\langle{(M^{-}_{x})}^4 \rangle} \\
    Q_s & ={\langle m_{s}^2\rangle}^2/{\langle{(m_{s}}^4 \rangle},
\end{align}
corresponding to the antiferro magnetization $M^{-}_{x}$ and the striped ferromagnetization 
$m_{s}$, respectively.

(\uppercase\expandafter{\romannumeral2})   Vortex density
\begin{align}
    \rho_{vx} & =\frac{1}{2f} \sum_{i=1}^{2f}   n _i
\end{align}
where $f$ is the \tcr{total number of faces} of one layer.
\\
\noindent (\uppercase\expandafter{\romannumeral 3})  Specific heats $C_V$,   and  $C_{vx}$ are expressed as:
\begin{align}
    C_V    & =\frac {N} {{k_B}T^2}[{\langle e^2 \rangle}-{\langle e \rangle}^2],                                                      \\
    C_{vx} & =\frac{N}{k_BT^2} \left [ \left \langle \rho_{vx}^2 \right \rangle -\left \langle \rho_{vx} \right \rangle ^2  \right ],
\end{align}
where $e$ is the average energy \tcr{per site}. $C_V$ and  $C_{vx}$ are specific heats related \tcr{to} energy and vortex.


\section{\texorpdfstring{phase diagram $\bf V=1$ ($\bf v=0$) and details}{}}
\label{sec:res1}

\subsection{\texorpdfstring{\tcr{Global} phase diagram with $ \bf V=1$}{}}

We first consider the global phase diagram with the vortex-excitation factor $V=1$. Traditionally, for a two-dimensional square lattice with periodic boundary conditions, the  variable $\Delta$ was introduced to describe the phase diagram~\cite{22in,22out},
\begin{equation}
    \begin{aligned} \Delta=\frac{a^2+b^2-c^2}{2ab},
    \end{aligned}
\end{equation}
and the four  phases, along with  their boundaries expressed in terms of the parameters $a$, $b$ , $c$ and $\Delta$, are listed in \tcr{Table}~ \ref{tab:my_label}.

\begin{table}[h]
    \centering
    \begin{tabular}{c|c|c|c|c}
        \hline
        \hline
        ~~~phase ~~~ & ~~~ $\Delta$~~~ & ~~~~~$~a,~b,~c~$~~~~ & $m^+_x$   & $m^{-}_x$ \\
        \hline
        ~~FM ~~      & ~~$\Delta>1$~~~ & $b - a > c$~~~       & ~~$1$~~~  & ~~~0~~~   \\
        \hline
        FM           & $\Delta>1 $     & $a - b > c$~~~       & ~~$1$~~~  & ~~~0~~~   \\
        \hline
        DIS (D)      & -1 $<\Delta<1$  & $a + b > c$~~~       & ~~$ 0$~~~ & ~~~0~~~   \\
        \hline
        AFM          & $\Delta<-1$     & ~~$a + b < c$ ~~~    & ~~$ 0$~~~ & ~~~$1$~~~ \\
        \hline
        \hline
    \end{tabular}
    \caption{The four  phases, and their boundaries in terms of the parameters $a$, $b$ , $c$ and $\Delta$.  }
    \label{tab:my_label}
\end{table}

\begin{figure}[t]
    {\includegraphics[width=0.45\textwidth,height=0.2\textwidth]{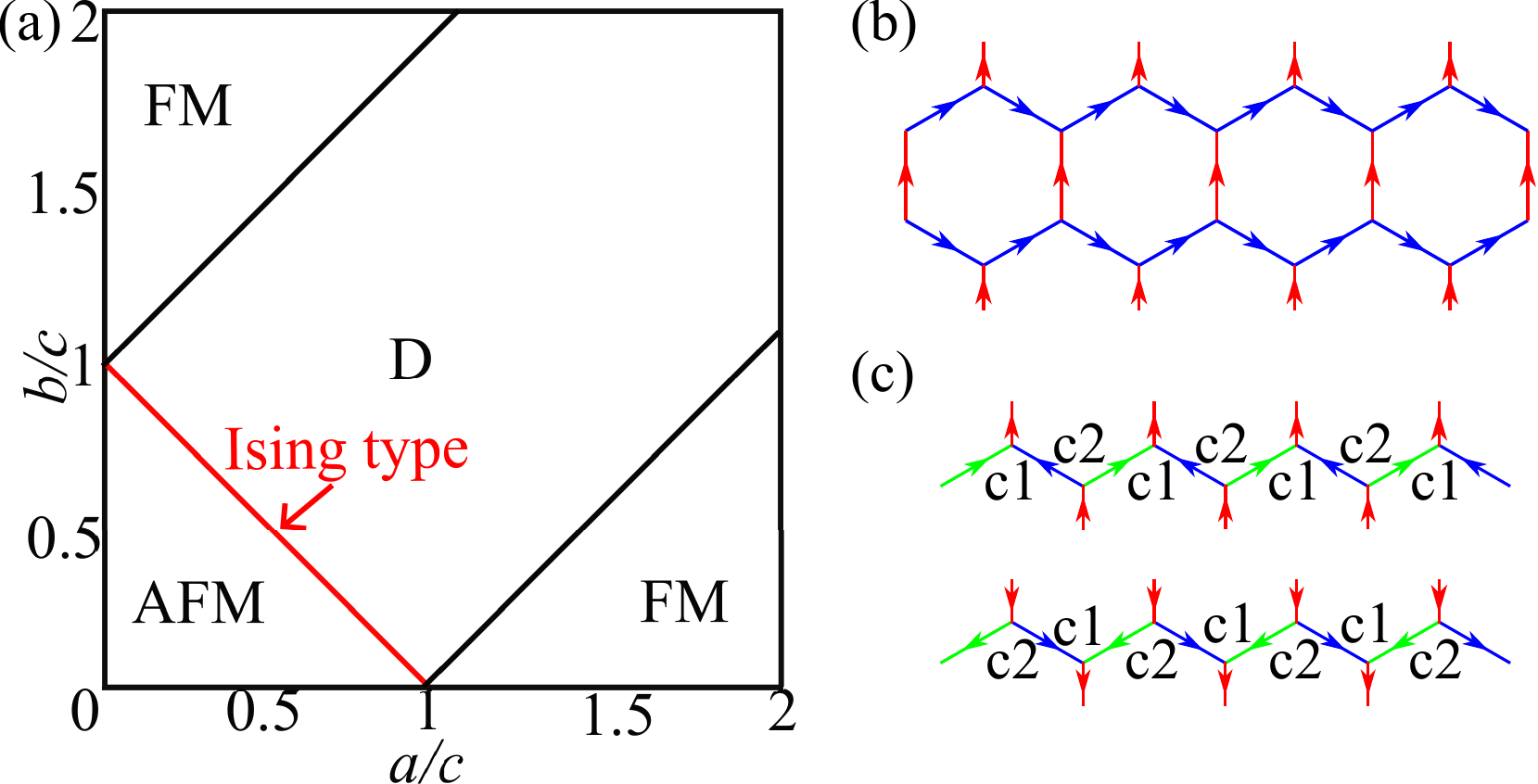}}
    \caption{ Global phase diagram in \tcr{the} plane ($a/c-b/c$) with $V=1$ and snapshots. (a) phase diagram containing AFM, FM and \tcr{disordered} phases (b) the typical snapshot in the FM phase. The AFM-D transition is of Ising type. (c) Two configurations of the AFM phase for the Z2 symmetry breaking.}
    \label{fig:phd}
\end{figure}

In Fig.~\ref{fig:phd} (a), the four phases are depicted, where the ferromagnetic (FM) phase, the antiferromagnetic (AFM) phase, and the disordered (D) phase are schematically located in Figs.~\ref{fig:phd} (b)-(d), respectively. For simplicity, we only show snapshots from the lower layer of the 6V model; the features of the other layer can be inferred based on symmetry.


Surprisingly, the phase diagram of the 6-vertex model on our bilayer honeycomb lattice is identical to the tabulated results in Table~\ref{tab:my_label} from the square lattice presented above~\cite{22in,22out}.

One may wonder why the two-layer 6V model is the same as the single-layer square lattice 6V model.
Let us now briefly analyze a few locations of the phase transition boundary. The first point is $(a/c,b/c)=(1,0)$ which is the point of phase transition between the AFM and FM phases. As illustrated in Fig.~\ref{fig:phd} (a), the FM-phase is full of vertices with weight $a$ and the AF-phase is full of vertices with weight $c$. At the point of phase transition of the two phases, i.e., phase AFM and phase FM, free energies are equal \tcr{and} defined \tcr{by the} following:


\begin{equation}
    N_a\text{ln}a-TS_{\text{FM}} =N_c\text{ln}c-TS_{\text{AFM}},
    \label{eq:eq}
\end{equation}
$N_a= N_c=2L^2$
where $S_{\text{AF}}$ and $S_{\text{FM}}$ are the entropies of the two phases, respectively,

\begin{equation}
    \begin{aligned}
       S_{\text{FM}}  & =S_{\text{AFM}}=k_B\text{ln}2. \\
    \end{aligned}
\end{equation}
 
Along the $b=0$ axis, the system in the FM phase  only has either all \tcr{a1 type} vertices or all \tcr{a2 type} vertices. In other words, there are two microscopic states.
Similarly, in the AFM phase, there exist configurations with only all c1 or all c2 type vertices.
Therefore $a/c=1$ is the critical point satisfying Eq.~\ref{eq:eq} at $b=0$.



\subsection{\texorpdfstring{Ising type not BKT type \tcr{in transition} along  $a=b$}{}}

The phase transitions  of the 6V model \tcr{are} identical on both the square lattice and \tcr{the} double-layer honeycomb lattice.
For example,
along $a=b$, Ref.~\cite{Fmodel}
confirms the critical point
is at  $\beta_c=\text{ln}(2)$.
The critical point locates at:
\begin{equation}
    a_c=b_c=1/2.
\end{equation}

In our MC simulation, as shown in Fig.~\ref{fig:v0} (a)-(b), \tcr{the} intersection of $m_x^{-}$ and  its Binder ratio $Q$
locates  at
$a_c=b_c=1/2$.

However, the type of the phase transition changes. In the $F$ model, the phase transition belongs to be the BKT type, as evidenced by the non-divergent behavior of specific heats in Fig.~4 in Ref.~\cite{Fmodel}.
In our bilayer 6V model,
the type should be \tcr{of} Ising type.

The first signal is that, for the Ising phase transition, the critical exponent $y_t=1$.  Using the finite size scaling equation defined as follows:
\begin{figure}[t]
    \includegraphics{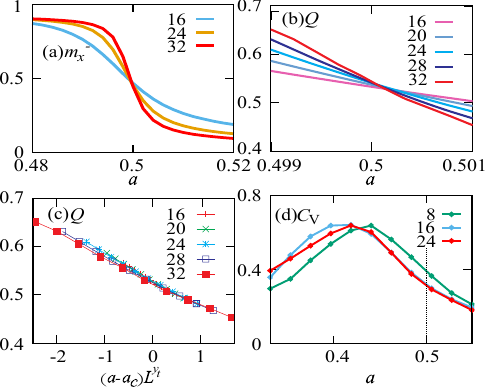}
    \vskip -0.4cm
    \caption{ Different quantities scanning  $a$ while keeping $a=b$ (a) $m_x^{-}$ (b) $Q$ (c) Data collapse of  $Q$, while using $y_t=1$. (d) Repeating $C_V$ of the $F$ model on the square lattice~\cite{Fmodel}.}
    \label{fig:v0}
\end{figure}

\begin{equation}
    \begin{aligned}
        Q= & Q_0+e_1(T-T_c)L^{y_t}+e_2(T-T_c)^2L^{2{y_t}} \\
           & +\cdots+f_1L^{y_1}+f_2L^{y_2}+\cdots.
    \end{aligned}
    \label{eq:q}
\end{equation}
and the data \tony{collapse} method, $Q$ is plotted as function of $(a-a_c)L^{y_t}$, and data of different sizes overlap as shown in Fig.~\ref{fig:v0} (c).
This phenomenon supports the conclusion that the observed phase transition is indeed the Ising type.

This type of Ising phase transition can also be analyzed from the perspective of symmetry breaking. In the AFM phase, as depicted in Fig.~\ref{fig:phd} (c), there is a twofold degeneracy in \tcr{the} configuration. In the horizontal direction, one configuration is characterized by a c1-c2-c1-c2 vertex arrangement, while the other features a c2-c1-c2-c1 arrangement. Moreover, the relationship between these two configurations is achieved by flipping the states of all legs.
This means that from disorder to AFM, there is a  Z2 symmetry breaking leading to the Ising transition.

To further \tcr{vindicate} our code, we also simulate the F model on the one-layer square lattice; the
$C_V$ indeed does not diverge at $a_c$ as shown in Fig.~\ref{fig:v0} (d), consistent with the result in Ref.~\cite{Fmodel}.


\begin{figure}[h]
    \centering
    \includegraphics[width=0.45\textwidth,height=0.4\textwidth]{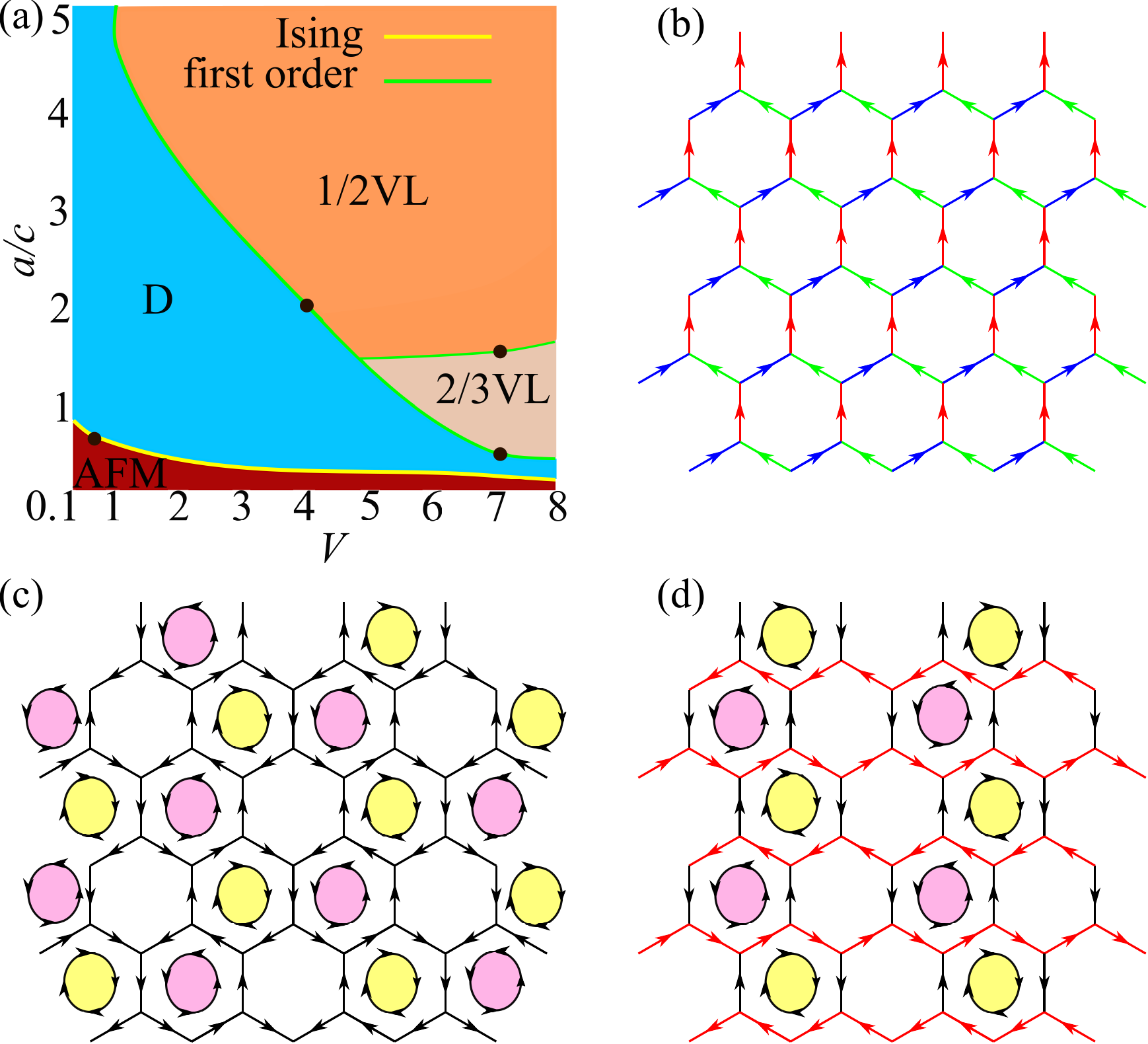}
    \caption{ Global phase diagram with $V\ne1$ and snapshots. (a) A schematic  phase diagram   (b) the AF phase (c) 2/3 VL phase (d)  1/2 VL phase. In fact, 1/2 VL is accompanied by \tcr{a} $x$-direction ferromagnetic order and 1/3 VL is accompanied by \tcr{an} AFM order (see appendix~\ref{sec:app}). }
    \label{fig:configuration}
\end{figure}
\section{\texorpdfstring{phase diagram with $\bf V\ne 1$}{}}
\label{sec:vortexne0}

In this section, we introduce a non-zero value for $v$ in Eq.~\ref{eq:hami} to discuss the \tcr{effects of} vortex excitation.

As in Ref.~\cite{ding}, it is experimentally possible to manually insert \tcr{or} delete  vortices~\cite{vortex-excitation1,vortex-excitation2}, despite the fact that the factors are added manually here.

This phase diagram is shown in the $a-V$ plane, where  $V=e^{-v}$, while simultaneously maintaining a fixed cut \tcr{along} $b=a$.

\subsection{\tcr{Global} phase diagram}

Figure \ref{fig:configuration} (a) shows the global \tcr{phase} diagram, which contains the AFM, \tcr{disordered}, 1/2  and 2/3 VL phases.  When the vortex excitation is considered, the phase diagram becomes rich \tcr{in features}.
In Fig.~\ref{fig:configuration} (c), the configuration of the 2/3-VL phase is shown.
The dual lattice of the honeycomb lattice is the triangular lattice. In this phase, two of the three sets of sublattices within the triangular lattice are occupied by positive and negative vortices, respectively, while the remaining set of sublattices contains no vortices.
In Fig.~\ref{fig:configuration} (d),  the configuration of the 1/2-VL phase is shown.  In this phase, positive and negative vortices alternate sequentially along the $y$-direction, forming connected stripes.

Here, we analyze the distribution of different phases within the phase diagram. When $a$ and $b$ are very small, $a=b\ll c$ and $V\ll c$, the system resides in the AFM phase, represented by the dark red region.
\tcr{In the} upper right region,
$a=b\gg c$ and $V\gg c$,
the system
corresponds to the 1/2 VL phase.
The system has almost no type $c$ vertices. At the same time $V$ drives the system to form vortices consisting of vertices of type $a$, $b$ only, as shown in Fig.~\ref{fig:snapshot} (a).

When $a = b \in  [0.5c, 2c]$, the influence of $V$ drives the system to form vortices consisting of vertices of types $a$, $b$, and $c$, maintaining a 1:1:1 ratio, and the typical configuration of 2/3 VL phase is  illustrated in Fig.~\ref{fig:snapshot} (b).

In the disordered phase, the characteristics of vortices vary across different parameter regimes. As depicted in Fig.~\ref{fig:snapshot} (e), two distinct regimes are identified, labeled as  $\textcircled{3}$ and   $\textcircled{4}$.

In the region where
$V<1$ and
$a(b)$ is large, as shown in Fig.~\ref{fig:snapshot} (c), vortices are absent. The reason is that a vortex typically requires the presence of vertices of types
$a$, $b$, and $c$,
within its structure.
Conversely, in the other regime where
$V>1$  and reasonable values of
$a(b)$ are considered, as illustrated in Fig.~\ref{fig:snapshot} (d), numerous vortices appear randomly throughout the system.

In fact, 1/2 VL is accompanied by \tcr{a} $x$-direction ferromagnetic order and 1/3 VL is accompanied by \tcr{an} AFM order (see appendix~\ref{sec:app}).

\subsection{Vortex structure factors}

To further understand the spin vortex lattice phase, \tcr{the} structure factor in $\bm{q}$-space is introduced as
\begin{equation}
    S(\bm{q})=\frac{1}{N_c} \sum_{i,j}^{} e^{i\bm{q}\cdot(\bm{r_i}-\bm{r_j}) }\left \langle n_in_j \right \rangle,
    \label{eq:sq}
\end{equation}
where $n_i$, $n_j$ represents 1, 0,  in the face of the honeycomb lattices, i.e., \tcr{a} triangular lattice. The symbols
$\bm{r_i}$ and $\bm{r_j}$ are the center coordinates of the vortex.
In real space, if the density  obeys configurations of the form  ($101010 \cdots$)  or  ($010101 \cdots$),
the wave vector corresponding to the maximum value of $S(\bm{q})$ should be located at  $\mathbf{q}=(\pi, 0)$~\cite{whh}.

\begin{figure}[ht]
    {\includegraphics[width=0.495\textwidth]{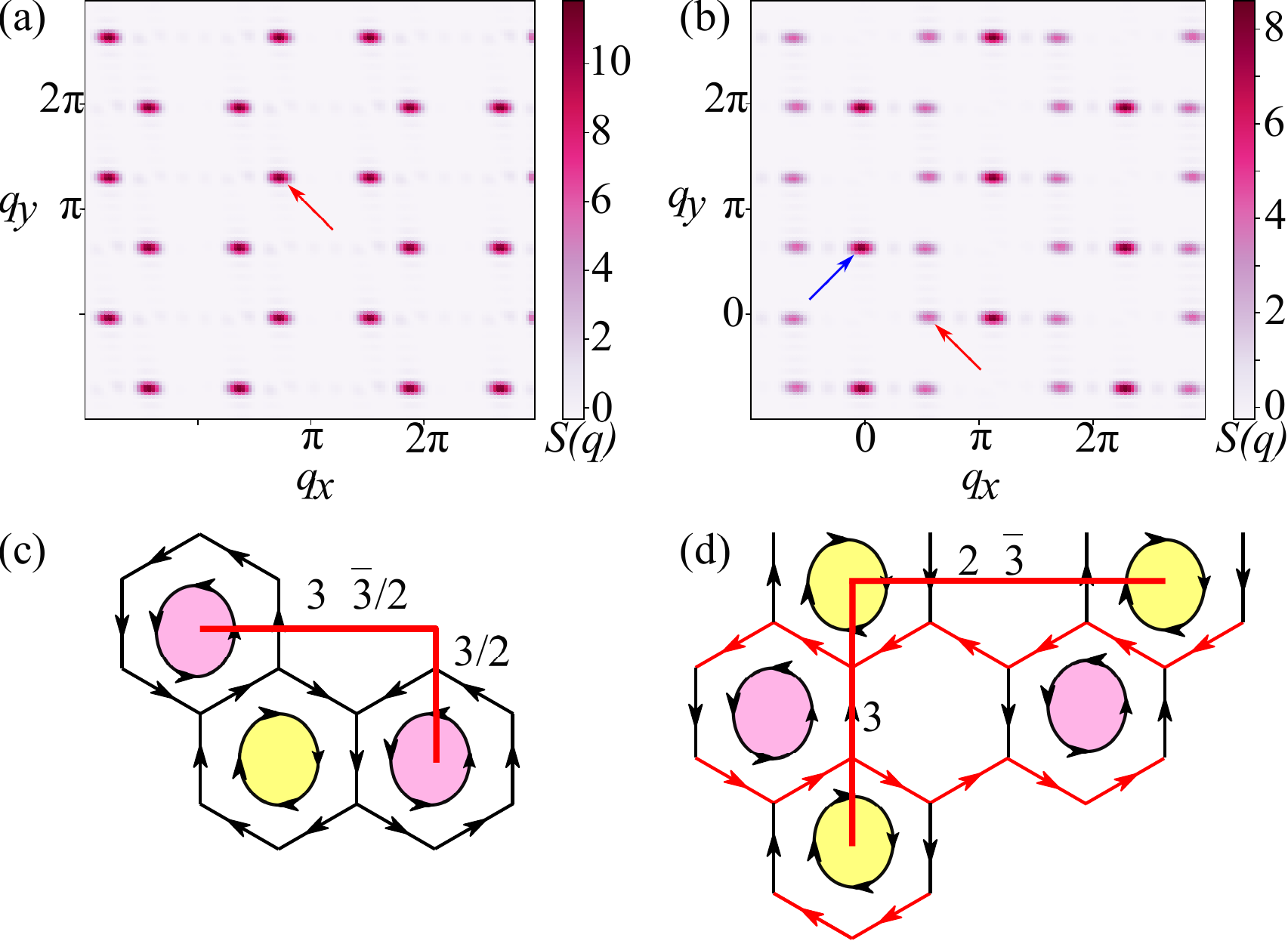}}
    \caption{ The vortex structure factor $S(\bold{q})$ defined in Eq.~\ref{eq:sq} for  (a)  2/3 VL phase (b) 1/2 VL phase in $\bold{q}$ space. The translation vectors in real space  (c)  2/3 VL phase (d) 1/2 VL phase.}

    \label{fig:sq1}
\end{figure}


In Fig.~\ref{fig:sq1} (a),
for the 2/3 VL phase, $S(q_x,q_y)$ is obtained by using Eq.~\ref{eq:sq} with a lattice size $L=12$. One of the brightest \tcr{points} is located at:
\begin{equation}
    (q_x,q_y)=(\frac{4\sqrt{3} }{9} \pi,\frac{4}{3} \pi ),
    \label{eq:qxqy}
\end{equation}
\tcr{as indicated} by the red arrow.
The position of the peaks reflects the translational symmetry of the vortex lattice.
Assuming that the side length of the honeycomb lattice is 1, the spacing between the two pink vortices $\bm{r_i}$ and $\bm{r_j}$  is:
\begin{equation}
    \Delta x=x_i- x_j= 3\sqrt{3}/2,
\end{equation}
and
\begin{equation}
    \Delta y=y_i- y_j=3/2.
\end{equation}
The \tcr{position} in Eq.~\ref{eq:qxqy} can be obtained by using:
\begin{equation}
    q_{x(y)}= 2\pi / \Delta x(y).
\end{equation}





For the 1/2 VL lattice, $S(q_x,q_y)$ is shown  in Fig.~\ref{fig:sq1} (b). The brightest point  (\tcr{as indicated} by the blue arrow)  is located at:
\begin{equation}
    (q_x,q_y)=(0,\frac{2}{3} \pi),
\end{equation}
and the second brightest point (\tcr{as indicated} by the red arrow) is located at:
\begin{equation}
    (q_x,q_y)=(\frac{\sqrt{3} }{3} \pi,0).
\end{equation}
This is because in real space, the translation vectors are respectively
$\Delta x=2\sqrt{3}$ and
$\Delta y=3$.

In total, we have identified two VL phases in the bilayer 6V model with vortex weight where
$V>1$. These ordered states can be distinguished by examining either their respective structure factors or the relevant order parameters.

\subsection{Size effects from periodic boundaries}

\begin{figure}[H]
    {\includegraphics[width=0.5\textwidth,height=0.4\textwidth]{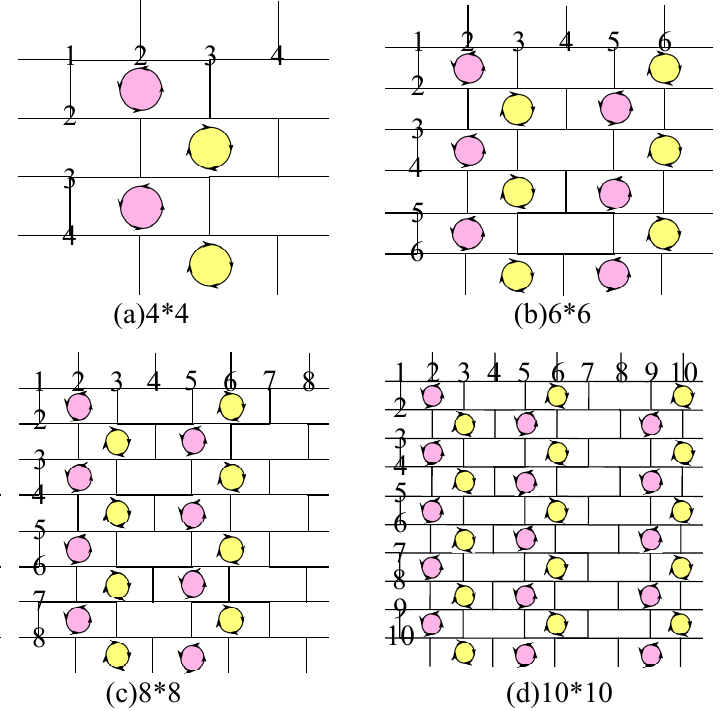}}

    \caption{Vortex and anti-vortex in the lattices with $L=4, 6, 8$ and 10. }

    \label{fig:enter-label}
\end{figure}

Figure~\ref{fig:enter-label} shows the size effects of the vortex density in a 2/3  VL with $L=4, 6, 8, 10$, respectively. The honeycomb lattice is a complex lattice with two sites in its smallest cell, hence having an even number of lattice sites in the horizontal direction, i.e., 
$L~mod~2 = 0$.

In Fig.~\ref{fig:enter-label} (a), for $L=4$, each row contains two faces, one of which is occupied by a vortex. The vortex density, i.e., the ratio between the number of vortices and faces, is $\rho_{v}=N_v/N_f=1/2$. Similarly, for $L=6, 8, 10$, the densities become $\rho_{v}=2/3$, 2/4, and 3/5, respectively, as shown in Figs.~\ref{fig:enter-label} (b)-(d). In Table~\ref{tab:rhov}, we list some possible small sizes and vortex densities.

\begin{table}[H]
    \caption{The table shows the maximum theoretical value of the vortex density $\rho_{v}$ as a function of system size $L$.}
    \centering
    \begin{tabular}
        {c|c|c|c|c|c|c|c|c|c|c|c}
        \toprule 
        phase  & $L$         & $4$   & $6$   & $8$   & $10$  & $12$  & $14$  & $16$  & $18$  & $20$   & $L$                                 \\ 
        \hline
        2/3 VL & $\rho_{vx}$ & $1/2$ & $2/3$ & $2/4$ & $3/5$ & $4/6$ & $4/7$ & $5/8$ & $6/9$ & $6/10$ & $\frac{\lfloor L/3 \rfloor}{L/2}$   \\
        \hline
        1/2 VL & $\rho_{vx}$ & $1/2$ & $1/3$ & $2/4$ & $2/5$ & $3/6$ & $3/7$ & $4/8$ & $4/9$ & $5/10$ & $~\frac{\lfloor L/4 \rfloor}{L/2}$~ \\
        \hline
    \end{tabular}
    \label{tab:rhov}
\end{table}

For more general sizes, the densities are as follows:

\begin{equation}
    \rho_v=\frac{\lfloor L/3 \rfloor}{L/2},
    \label{eq:rho}
\end{equation}
where the symbol $\lfloor . \rfloor$ denotes rounding down.
We first explain Eq.~\ref{eq:rho} using $L=6$, two sites labeled as 1,~2 in 
\tcr{the} horizontal direction, a total of $L/2$ bricks in each row as shown in
\tcr{the} denominator.

Then we explain the numerator in Eq.~\ref{eq:rho}.
The following three sub-equations explain the number of vortices corresponding to the three sizes $L$, $L+2$, and $L+4$, as shown below:
\sublabon{equation}
\begin{eqnarray}
    \frac{L}{2} \times \frac{2}{3}&=&\frac{L}{3}  \rightarrow\frac{L}{3}=\lfloor \frac{L}{3}\rfloor,\\
    \frac{L+2}{2} \times \frac{2}{3}&=&\frac{L}{3}+  \frac{2}{3} \rightarrow  \frac{L}{3}=\lfloor \frac{L+2}{3}\rfloor, \label{eq:l2}\\
    \frac{L+4}{2} \times \frac{2}{3}&=&\frac{L}{3}+\frac{4}{3} \rightarrow  \frac{L}{3}+1=\lfloor \frac{L+4}{3}\rfloor,
\end{eqnarray}
\sublaboff{equation}
\noindent where $L+2$ in Eq.~(\ref{eq:l2}) indicates two additional sites compared to $L$, i.e., a new empty face without a vortex ($n_i=0$), as illustrated by comparing Figs.~\ref{fig:enter-label} (b) and (c) in the first lines.

Through the analysis above and the densities presented in Table~\ref{tab:rhov}, in the 2/3 VL phase, the density remains constant for sizes where 
$L~mod~6 = 0$. 
Therefore, when observing physical quantities later \tcr{on}, we only simulate systems whose size is a multiple of 6.

\subsection{Detailed transitions between the phases}
In this section, our focus is on examining the specific details of phase transitions between multiple phases, and analyzing its underlying reasons.

\subsubsection{\texorpdfstring{The Ising  transition between the AFM and disordered phases at $V<1$}{}}

\begin{figure}[h]
    \centering
    \vskip -0.3cm
    \includegraphics{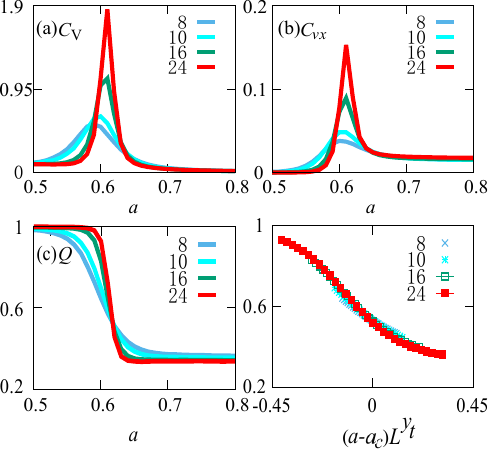}
    \vskip -0.3cm
    \caption{ The Ising  transition between the AFM and \tcr{disordered} phases at $V<1$, $a=b$ and  $V=0.5$.
        (a) $C_V$ (b) $C_{vx}$, (c) $Q$ (d) data collapse of $Q$, using $y_t=1$. The system size  ranges from 8 to 24.}
    \label{fig:v=0.5}
\end{figure}

Initially, we scan the parameter  $a=b \in [0.5,0.8]$ at $V=0.5$,  around the black point in the lower left corner of  the phase diagram.
Various  quantities are shown in Fig.~\ref{fig:v=0.5}.


In Figs.~\ref{fig:v=0.5} (a) and (b), the specific heat $C_V$ and $C_{vx}$  exhibit divergence with respect to sizes during the phase transition between the AFM and \tcr{disordered} phases. This behavior stands in contrast to the {\it convergence} of $C_V$ in the single-layer 6V $F$ model~\cite{Fmodel}, as shown in Fig.~\ref{fig:v0} (d).

In Figs.~\ref{fig:v=0.5} (c) and (d), the  Binder cumulant $Q$, along with its data collapse at the critical point $a_c=0.617$, yields $y_t = 1$, providing additional confirmation of the Ising transitions.

%
The possible reasons of the Ising  transition between the AFM and disordered phases at $V<1$. Be precise are as follows: $V<1$ implies the absence of the disordered phase as depicted by the absence of vortices in Fig.~\ref{fig:snapshot} (c). Additionally, the AFM phase has no  vertex, as illustrated in Fig.~\ref{fig:configuration} (d). Consequently, the phase transition between these two phases does not involve vortices. Moreover, the transition from simple AFM \tcr{order to disorder} involves $Z2$ symmetry breaking.


\subsubsection{\texorpdfstring{
        \tcr{Transitions} between the D and 1/2 VL phases at $V>1$}{}}

The phase transition between \tcr{disordered and the} 1/2 VL phase is discussed.
By fixing the parameter $V=4$, and scanning $a$ and keeping $a=b$, the vortex density $\rho_{vx}$, and $C_V$
are measured for different sizes $L=16-48$.

\begin{figure}[t]
    \centering

    \includegraphics[width=0.48\textwidth,height=0.6\textwidth]{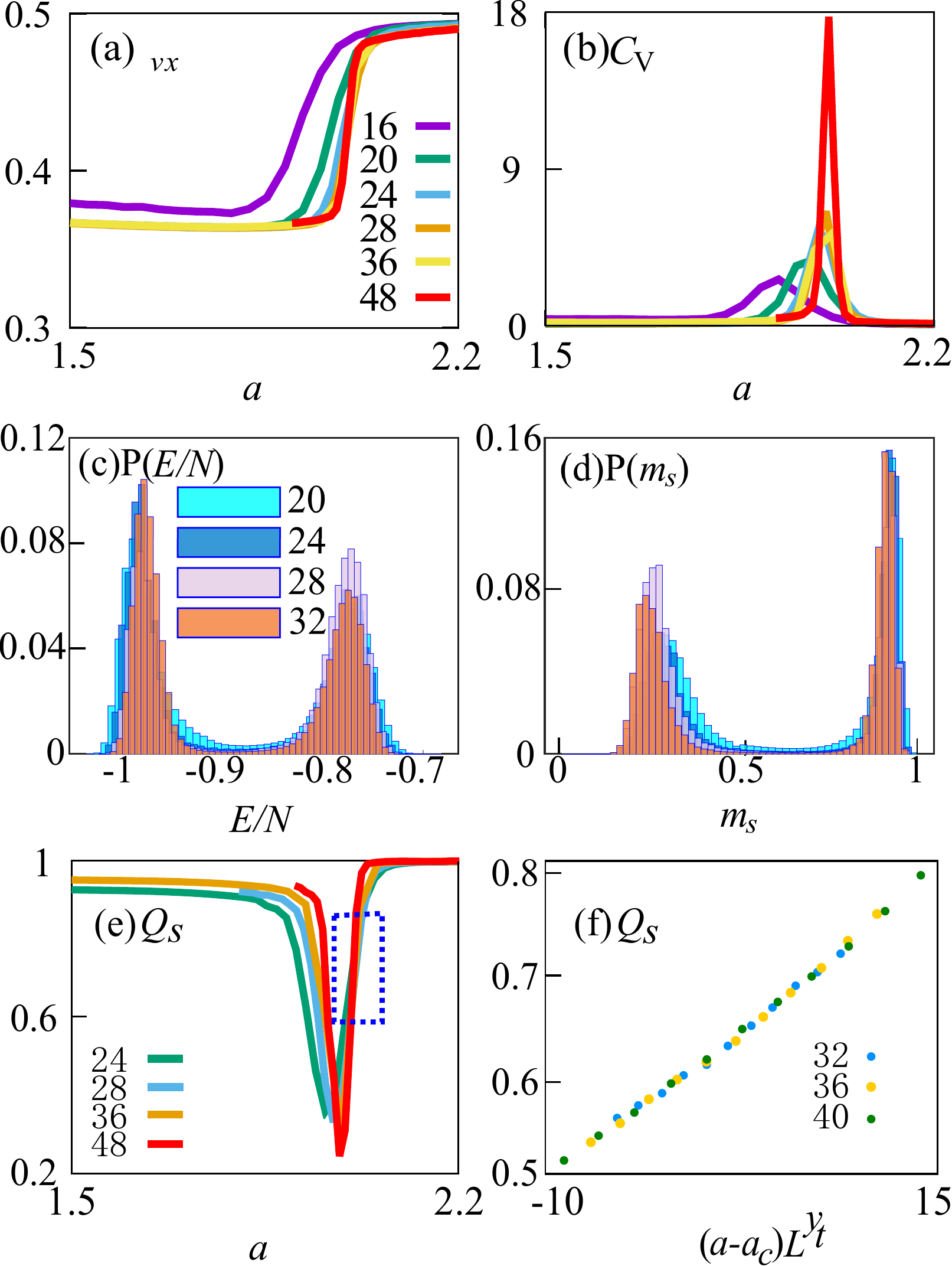}
    \vskip -0.3cm
    \caption{The signatures of the first order phase transition at $V=4$. The quantities (a) $\rho_{vx}$  (b) $C_{V}$   with system sizes $L=16-48$.  The
        double peaked distributions (c) $P(E/N)$ and (d) $P(m_s)$ at  $a=2.004(1)$. (e)  $Q_s$ and (f) data collapse of $Q_s$ and obtaining $\nu=0.48(8)$~\cite{nihe}.}
    \label{fig:v=4}
\end{figure}

As shown in \tcr{Figs.}~\ref{fig:v=4} (a) and (b), the jumps of  $\rho_{vx}$ and peaks of  $C_V$ show the signatures of phase transitions.
The  divergence of $C_V$  confirms the phase transition is not \tcr{of} BKT type~\cite{ding}.
In Figs.~\ref{fig:v=4} (c)-(d), the \tcr{histograms} of $E/N$ and $m_s$ are 
\tcr{obviously double peaked} at the parameters $a=b=2$. 
This indicates the phase transition 
\tcr{is of first order}.

The signature of the first order transition can also obtained by fitting the critical exponent.
Using the package for finite size scaling~\cite{nihe}, $\nu$ is obtained as $0.48(8)$ using Eq.~\ref{eq:q}.

The data for sizes $L=32, 36, 40$
\tcr{overlap} very well.
The scaling dimension $y_t=1/0.48=2.08\approx 2$ equals to the system dimension $d=2$ when the first order transition occurs~\cite{ziyangwang}.

\subsubsection{\tcr{Transition} between the D and 2/3 VL phases }

\begin{figure}[t]
    \centering
    \includegraphics[width=0.5\textwidth,height=0.59\textwidth]{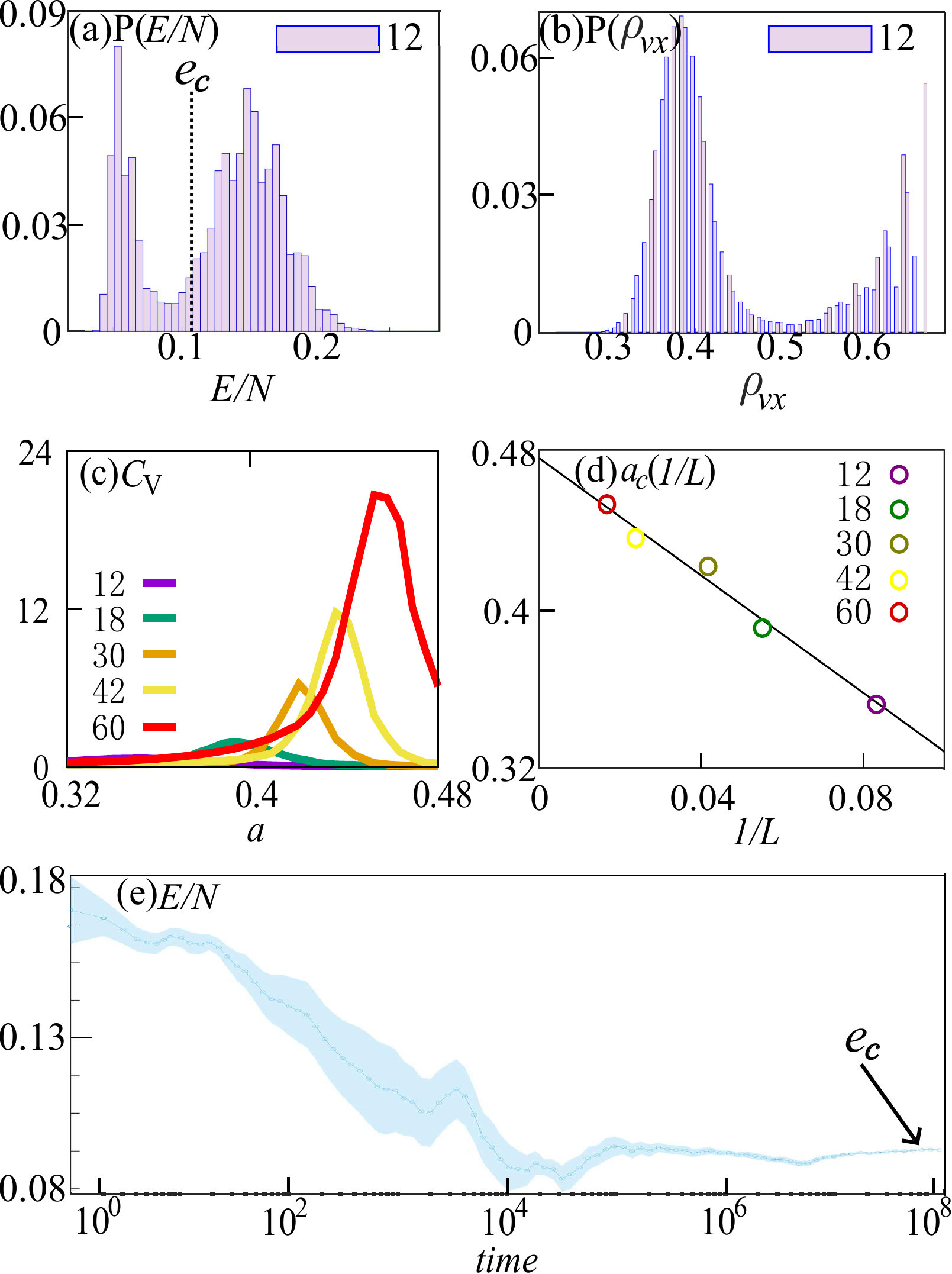}
    \vskip -0.1 cm
    \caption{The signatures of the first order phase transition. The
        double peaked \tcr{distributions}  of (a) $P(E/N)$ and (b) $P(\rho_{vx})$ at $V=7$ and $a=0.35$. (c) $C_V$ versus $a$, (d) $a_c$ versus $1/L$. (e) $E/N$ versus MC steps and the convergent time is about $10^6$. }
    \label{fig:v=7}
\end{figure}
The transition from the 2/3 VL phase to the disorder phase is also first order.
In Figs.~\ref{fig:v=7} (a) and (b), the double peaks in the distribution of $P(E)$ and $P(\rho_{vx})$ indicating a first order phase transition.

Different from \tcr{the} 1/2 VL phase as shown in Fig.~\ref{fig:v=4}, this phase transition has an obvious size effect.
Figure~\ref{fig:v=7}(c) illustrates that the specific heat peak shifts to the right as the size increases. Fortunately, through finite-size scaling defined as follows:
\begin{equation}
    \begin{aligned}
        a_{c} (1/L)=K/L+a_{c} \left ( L=\infty  \right ),
    \end{aligned}
    \label{fss}
\end{equation}
the position of the specific heat converges under the thermodynamic limit $a_{c} \left (L= \infty  \right )=0.475811(2)$.
The line $a_{c} (1/L)$ versus $1/L$ is
shown in Fig.~\ref{fig:v=7} (d).

The error bar (2) from 0.475811(2) is calculated
by the following equations.
First,  if one fits $y = kx+b$, then the standard deviation of the intercept b is
\begin{equation}
    s_b=\sqrt{\frac{n}{n\sum_{i}^{n}x_i^2-(\sum_{i}^{n}x_i )^2    } } \times s_y
\end{equation}
where $n$ is the number of points involved, $s_y$  is the standard deviation of the observation $y_i$ and can be expressed as:
\begin{equation}
    s_y=\sqrt{\frac{\sum_{i}^{n}(y_i-kx_i-b)^2}{n-2} }
\end{equation} where
$n-2$ is the degree of freedom.
If one fits the data using \tcr{``gnuplot'' software},  the result is
$a_{c} \left (\infty  \right )=0.476(5)$ consistent with the above result within the \tcr{error bar}.

To show that the results are convergent, Fig.~\ref{fig:v=7} (e) displays $E/N$ versus MC steps. The converging time is about $10^6$  steps. The converged energy  is approximately 0.12, \tcr{and} denoted as $e_c$ \tcr{which} is also marked by a dashed line in Fig.~\ref{fig:v=7} (a), sitting between the two peaks.

\subsubsection{First-order transition between the two vortex lattice phases}

\begin{figure}[h]
    \centering
    \includegraphics[width=0.5\textwidth,height=0.4\textwidth]{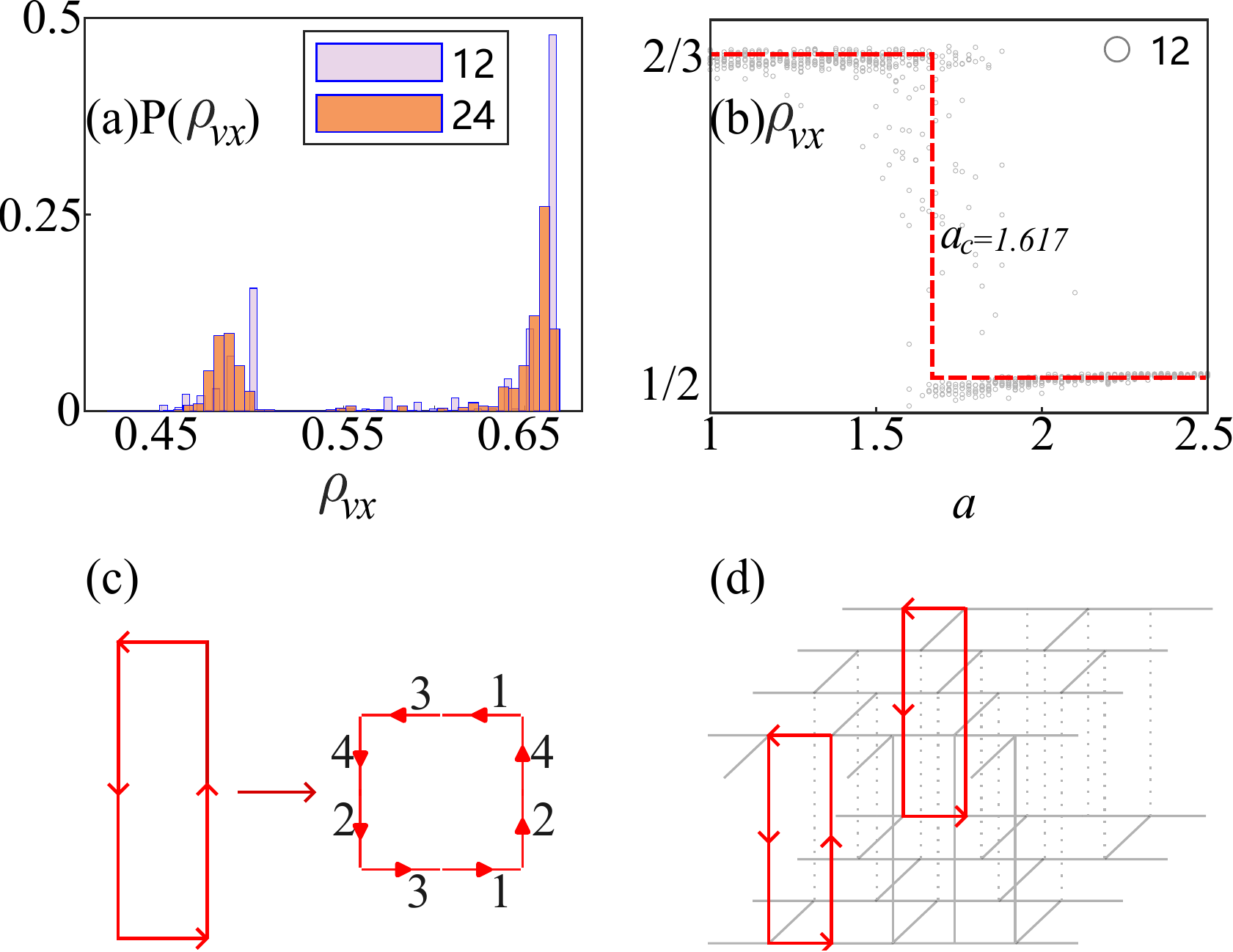}
    \caption{Signal   of the first order phase transition. (a) The 
\tcr{double peaked distribution} of  $\rho_{vx}$, with $V =7$, $a=1.68$ (b) $\rho_{vx}$ versus $a$ (c) a short loop in the vertical direction,  and the numbers 1, 2, 3, 4 represent the legs. (d) The \tcr{Metropolis type short-loop} updating.}
    \label{fig:pxhh}
\end{figure}

Similar to the atomic solid phase in the classical limit~\cite{Lilaixi.134536,zwz1,Liran,yinruoxi} for the BH model, the phase transition between 1/2 VL and 2/3 VL phases, should be \tcr{of first order}. The exact boundary between these two phases can be obtained analytically by comparing the free energy $F = E - TS$ \tcr{of both} phases.

By convention, the temperature is set to be $T=1$. By \tcr{careful checking,} the entropy for \tcr{both phases is respectively} $S_{1/2}=\ln_{}{64} $ and $S_{2/3}=\ln_{}{36}$. The entropy does not depend on the lattice size, and therefore, the average entropy for \tcr{per site}  denoted as $S/L^2/2$ should be zero in the thermal dynamical limit $L\rightarrow\infty$.
The energies of the two phases are defined as follows:
\begin{equation*}
    \left\{
    \begin{array}{lc}
        %
        E_{1/2\text{VL}}=-L^2\ln_{}{a}-L^2\ln_{}{b} -\frac{1}{2}L^2\ln_{}{V}, \\

        E_{2/3\text{VL}}=-\frac{2}{3} L^2\ln_{}{a}-\frac{2}{3} L^2\ln_{}{b}-\frac{2}{3} L^2\ln_{}{c}-\frac{2}{3}L^2\ln_{}{V}
    \end{array}
    \right.
\end{equation*}
where  $- L^{2}\ln_{}{a}- L^{2}\ln_{}{b}$  and $-\frac{2}{3} L^{2} \ln_{}{a}-\frac{2}{3} L^{2} \ln_{}{b}-\frac{2}{3} L^{2} \ln_{}{c}$ are the \tcr{vertex 
energies} for both phases, and $-\frac{1}{2}L^{2}  \ln_{}{V}$  and $ -\frac{2}{3}L^{2} \ln_{}{V}$ are
the energies for the vortex. Let
\begin{equation}
    F_{1/2 \text{VL}}=F_{2/3 \text{VL}},
\end{equation}
\tcr{while} keeping $a=b$, the reduced analytical expression becomes:
\begin{equation}
    a^4=V,
\end{equation}
when $L = \infty$.

\tcr{Numerical} simulations \tcr{are also} performed to confirm the first order transition between the vertex phases.
In Fig.~\ref{fig:pxhh} (a),
the \tcr{distributions} of $\rho_{vx}$ for the sizes $L=12, 24$ are obtained.
In Fig.~\ref{fig:pxhh} (b),
\tcr{the curves} $\rho_{vx}$ versus $L$ are plotted. The gray data are MC results with 20 independent bins.
The theoretical $a_c$,  are marked by the red line to guide the eyes.

In the regime of a strong first order transition, the cluster algorithm encounters the issue of ergodicity. As a result, we have also incorporated the \tcr{Metropolis type short-loop} update scheme. The short loop in the $z$-direction is composed of four vertices, with each vertex contributing two legs that form the rectangles depicted in Figure 11(c). These pairs of legs are (3,4), (2,3), (1,2) and (4,1), respectively.
In order to keep the \tcr{``two-in-two-out''} constraint, we flip eight legs simultaneously with a certain probability without violating the ``ice" rule for each involved vertex. 
\tcr{During} the actual simulation, after each round of loop updates, \tcr{we check} if there are any short loops present. If a short loop is found, we attempt to flip it.
\subsubsection{Effect of temperature gradient field}


\begin{figure}[H]
    \centering
\includegraphics[width=0.48\textwidth]{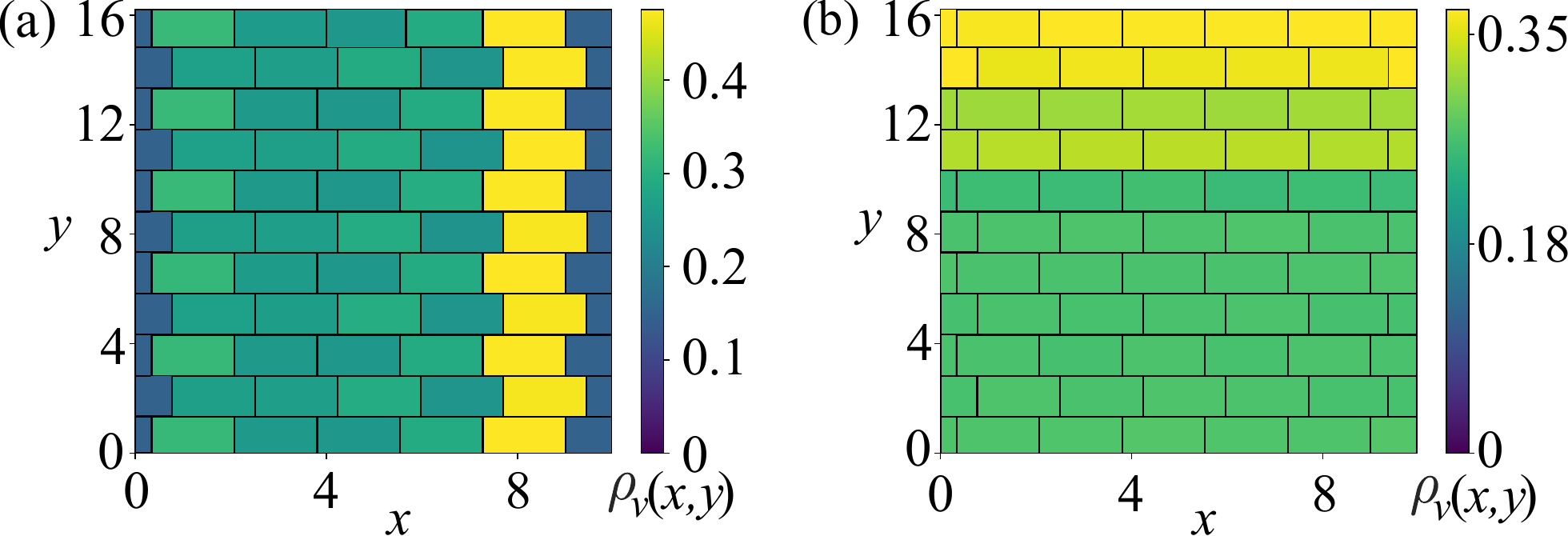}
    \caption{The distribution of vortex density $\rho_{v} (x,y)$ with temperature gradient \tcr{field} (a) along the $x$ direction  and (b) along the $y$ direction. The temperature varies from 1 to 0.45 with an interval of 0.05, with $a=b=1.5, c=1, V=2$ at the site (1,1).}
    \label{fig:td}
\end{figure}
On the single-layer honeycomb lattice, vortex lattice has been  achieved using the charged colloidal particle-based ice numerically~\cite{cj2}. The Coulomb repulsion between  charged particles favors  vortices at low temperatures.
Experimentally, \tcr{a} magnetic colloidal particle~\cite{ex_mag} inside lithographically
sculptured double well traps,
with tunable   repulsive interactions of  magnetic moments, also favors vortices.

However, raising the temperature ~\cite{cj2} results in the vortex state melting into a disordered state. Additionally, introducing a magnetic field~\cite{field} or an electric field in the plane also easily transitions the vortex state into a ferromagnetic state.


\tcr{Another factor of the vortex depinning is  a temperature gradient field.
Here we put a temperature field distributed on the lattice, with the temperature decreasing. The pattern of vortex density $\rho_v(x,y)$ becomes  striped as shown in Fig.~\ref{fig:td} (a)-(b). The vortex density $\rho_v(x,y)$ at higher temperature region are larger than that in the  lower temperature region.
Specifically, the coordinate variables
$x_1,x_2,\cdots, x_L$ are mapped to
$T_1,T_2,\cdots, T_L$, and ultimately to
$a_1,a_2,\cdots,a_L$ by $a_i=b_i=\text{exp}(\epsilon_i /T_i)$.
\tcr{A similar mapping applies to} the vortex weight $V_i$.}

Note that before applying the temperature gradient field, we set the temperature to 1. Then, we calculate the energy of each vertex in the lattice based on parameters $a$, $b$, $c$ and the Boltzmann weight. When introducing the temperature gradient field, it's crucial to maintain these energies fixed, as vertices at different positions experience different temperatures.

When an external field is applied, vortices either disappear or form stripes. Without an external field, due to translational invariance, the vortex density is uniformly distributed. The different average distributions of vortex density with and without an external field indicate vortex depinning.

\section{Experiment realization using particle based ice}
\label{sec:ex}

The colloidal particle-based ice, as discussed in \cite{cj1}, is another system capable of realizing bilayer ice, in addition to water ice, natural spin ice, and artificial spin ice. The theoretical concept behind colloidal particle ice involves employing elongated optical traps to confine colloidal particles. Each optical trap features two wells where colloidal particles can reside. These arrays of traps can organize into different kinds of lattices~\cite{cj1, cj2, cj8}. Additionally, the presence of topological charges~\cite{cj3}, vortex states~\cite{num1}  and skyrmions patterns~\cite{cj4}, boundary effects~\cite{boundary},  can be simulated by  particle-based ice.  More complete reviews
are  present in Ref.~\cite{cj7}.

\begin{figure}[H]
    \centering
    \includegraphics[width=0.42\textwidth]{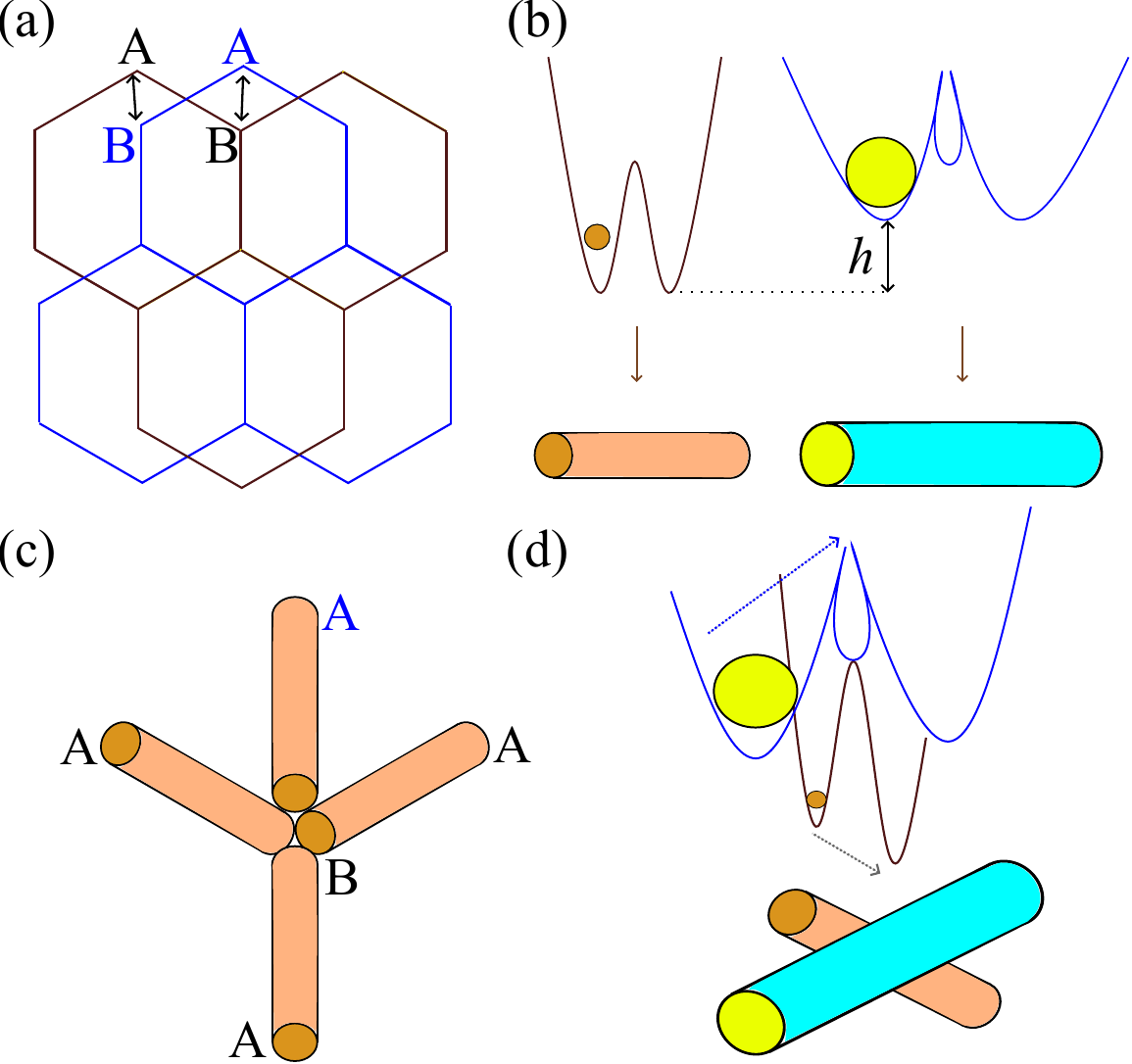}
    \vskip -0.3cm
    \caption{The method of embedding the A-B stacked bilayer honeycomb lattice into the single-layer substrate (a) The single-layer rendition of two honeycomb lattices, with double-sided arrows indicating their connection. (b) Representation of two different double well traps for the two lattices, each accommodating a single colloidal particle. These traps can be symbolized as bonds. (c) Illustration of the ``two-in-two-out'' rule pattern, where site B connects four \tcr{A sites}, including one A site from a different lattice. (d) Depiction of crossing bonds and their corresponding double well structures. }
    \label{fig:ex}
\end{figure}

Experimentally,
bilayer water
ice, where the ordering is due to the proton positions~\cite{jiangying}, is probably better captured by particle-based ice~\cite{cj7}.
Particle-based systems could include charged
or magnetic colloids interlacing in an array that mimics spin ice and
more closely resembles the water ice's proton ordering  despite their different \tcr{natures}~\cite{field}. The implementation of a two-well trap array embedded in a substrate has been achieved in references \cite{cj5,cj6,ex_mag},  on two-dimensional lattices. However, for a bilayer lattice, achieving the required structure has not yet been accomplished.

Here, we propose imprinting both layers onto a single-layer substrate, as schematically illustrated in Fig.~\ref{fig:ex} (a). The upper and lower honeycomb lattices are depicted in blue and black, respectively. Interlayer connections are denoted by double arrows. Given that this is an A-B stacked bilayer, the sites on the \tcr{A sublattice} and the \tcr{B sublattice} are connected pairwise.

As depicted in Fig.~\ref{fig:ex} (b), each edge can be represented as a double well  holding only one colloid. Some edges corresponding to the two  layers intersect, raising the possibility of particle collision at these \tcr{intersections}. To reduce this, the depths of the two wells can be adjusted differently. The black wells are deep, while the blue ones are shallow. Additionally, a small well between the two blue wells helps the passage of small colloids.


In Fig.~\ref{fig:ex} (c), the arrangement illustrates four non-crossing bonds. Site B (in the lower layer) connects to four A sites, with the blue A site originating from the upper layer. In Fig.~\ref{fig:ex} (d), the crossing bonds are depicted. The shallow and deep wells meet at the intersection from different directions. Both the big and small colloids can move  within their respective wells.





\section{ Discussion and conclusion}
\label{sec:conclusion}

In this paper, we model two-dimensional two-layer water ice as a two-layer 6V model.
By means of the loop update Monte Carlo method, we obtain interesting results.

Our study contrasts the square lattice 6V model, where the AFM to \tcr{the} disordered phase transition is governed by the BKT mechanism, with the bilayer 6V model, characterized by a conventional Ising phase transition due to Z2 symmetry breaking. We find that the transitions from vortex lattice phases to the disorder phase and between different vortex lattice phases are both of first order. \tcr{This insight advances} our understanding of the phase transitions present in layered water ices, contributing to the broader comprehension of complex systems in physics.

Despite conducting numerous simulations, there are still many open questions.

(i) As an initial investigation, we assumed $a=b$, and the case where $a\ne b$ has not been studied yet. The exploration of vortex glass induced by random values of $a$, $b$, and $V$ worthwhile. Furthermore, the disorder we observed can potentially be further classified into disordered structures with vortices and disordered structures without vortices.

(ii) For the bilayer water ice ~\cite{jiangying}, the systems  have various types of boundaries, such as  zig-zag and armchair edges, and rough random edges. It is interesting to simulate the bilayer honeycomb 6V model with different boundaries.

(iii) Although we are not clear about how water ice regulates the ratios of $a$, $b$, and $c$ types of \tcr{vertices}, it is possible to artificially adjust them in the case of artificial spin ice~\cite{arm}, 
which \tcr{is rich in physical properties}.

(iv) Regarding the numerical methods for studying this model, \tcr{not only are conventional MC methods suitable}, but tensor network methods~\cite{trg} are also well-suited for exploring this model. There are existing literature studies that have employed tensor network methods to investigate similar models such as the dimer model and \tcr{the} ice model. However, in our case, we introduce a slightly more complex factor by incorporating vortex weighting, which results in an increase in the bond dimensions of the tensor elements.


\vskip 0.5 cm
    {\it \tcr{Acknowledgments}}
We thank  Vladimir Korepin and Tzu-Chieh Wei for their inspiration  regarding the six-vertex model, and  Youjin Deng,  Chengxiang Ding, 
\tcr{and tcr C. Scott of RWTH-Aachen University} 
for commenting \tcr{on} the manuscript, and valuable suggestions from the reviewers. 
This work  was supported by the
Hefei National Research Center for Physical Sciences at
the Microscale (KF2021002), and \tcr{the} Shanxi Province Science Foundation (Grants No: 202303021221029 and No:202103021224051).

\appendix




\appendix
\section{Configurations with the $a$, $b$, $c$ types \tcr{vertices} and \tcr{vortices}}
\label{sec:app}
In Figs.~\ref{fig:snapshot} (a)-(d), the snapshots of the configurations 
with \tcr{vertices} and \tcr{vortices} are shown.
The parameters chosen correspond to the positions marked as \ding{172}, \ding{173}, \ding{174}, \ding{175} in  Fig.~\ref{fig:snapshot} (e).

In the 2/3 vortex lattice phase, \tcr{not only do vortex orders exist} but also AFM orders with a length of 3 in the $x$-direction Fig.~\ref{fig:snapshot} (f).
If $y==1$, $S_{y,x}=1$ for $x \in [1,2,3]$ and
$S_{y,x}=-1$ for $x \in [1,2,3]$.
The definition to quantify such \tcr{a} pattern is as follows:
\begin{align}
    M_3=\sum_{i=1}^{L}\left | \sum_{j=1}^{L} (-1)^{\left \lfloor  (j-1)/3 \right \rfloor} S_{i,j}^x  \right | /N,
\end{align}
where $\lfloor  (j-1)/3  \rfloor $ means replacing 1,2,3,4,5 and 6 to 0,0,0,1,1, and 1.
\tcr{Similarly}, for the 1/2 VL lattice, there is \tcr{a} FM order in the $x$ direction and the \tcr{quantity} is defined in Eq.~\ref{eq:ms}.

\section{Description of the bounce probabilities}

\setcounter{figure}{0}
\renewcommand{\thefigure}{A\arabic{figure}}

\begin{figure}[h]
    \includegraphics[width=0.48\textwidth]{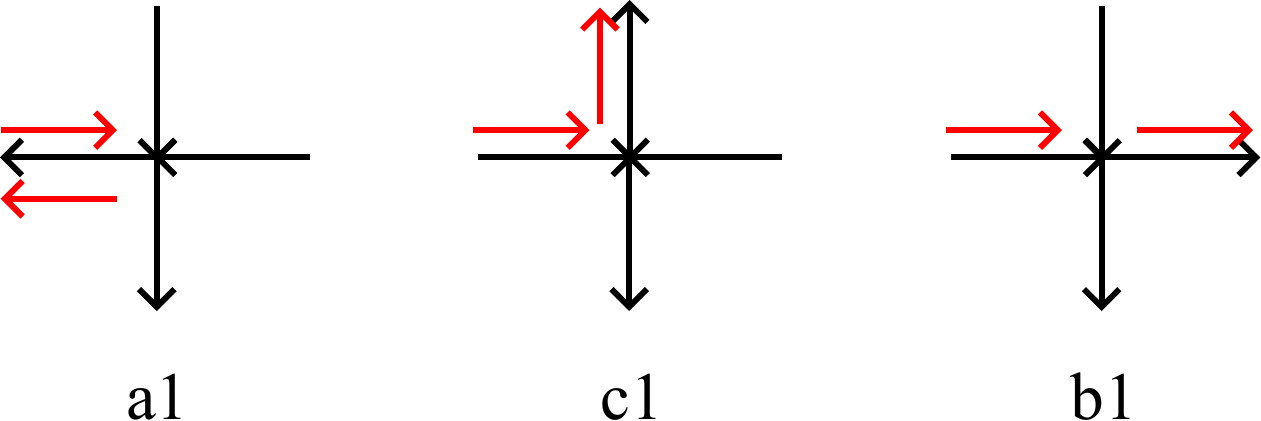}
    \caption{Two allowable exit legs and one bounce leg for the 6V model.}
    \label{fig:fantan}
\end{figure}
In Sec.~\ref{sec:algorithm}, Table~\ref{tab:my_label1} shows the range of bounce probabilities $p_a$, $p_b$ and $p_c$.
In this appendix, more details are described. Corresponding to the 6V model, the general expression of the bounce probability of the loop algorithm is:
\begin{equation}
    p_i=\frac{w_i}{w_a+w_b+w_c},
    \label{eq:pi}
\end{equation}
where the denominator is the sum of the weights of the three vertices after the loop passes through two allowable exit legs and one bounce leg (Fig.~\ref{fig:fantan}), and the numerator $w_i$ \tcr{are} the weights of the vertices encountered during the loop walk.  The subscript $i$ denotes types of $a$, $b$, or $c$. To discuss the range of $p_i$, it is only necessary to discuss the maximum and minimum values of Eq.~\ref{eq:pi}, where the variable $w_c$ is already fixed to 1.
Taking $p_a\in (0, 1/2)$ in the AFM phase as an example, the weights $w_a$ and  $w_b$ are independently adjustable, but they meet the range  $ 0 < a + b < 1$, and therefore the \tcr{maximum value} of $p_{a} = a/(a + b + 1) \rightarrow a/(a + 1)\rightarrow 1-1 / (a + 1) = 1/2$.


\begin{figure*}[p]
    \centering
    \includegraphics[width=0.48\textwidth,height=0.52\textwidth]{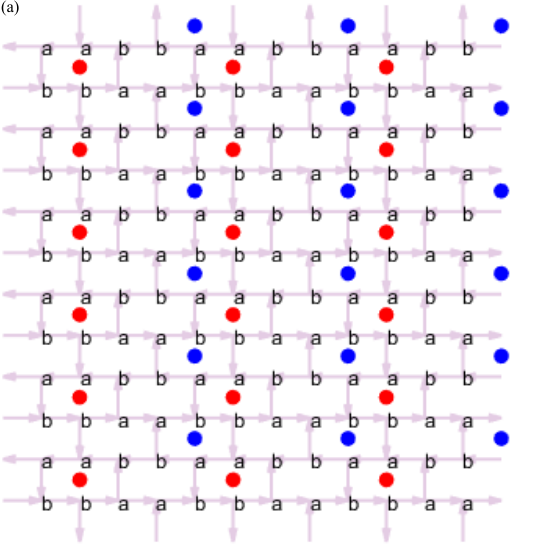}
    \includegraphics[width=0.48\textwidth,height=0.52\textwidth]{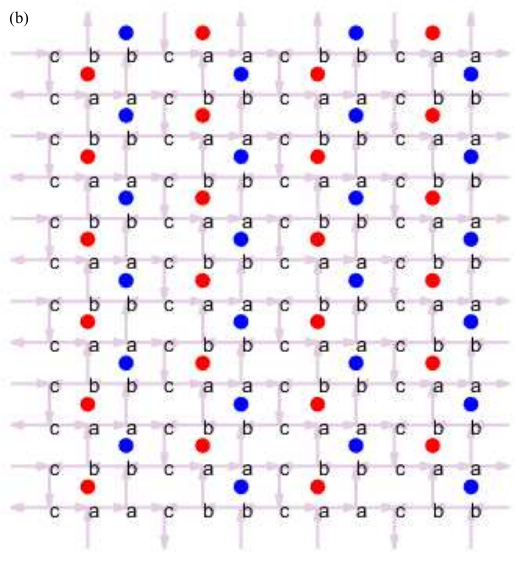}
    \includegraphics[width=0.48\textwidth,height=0.52\textwidth]{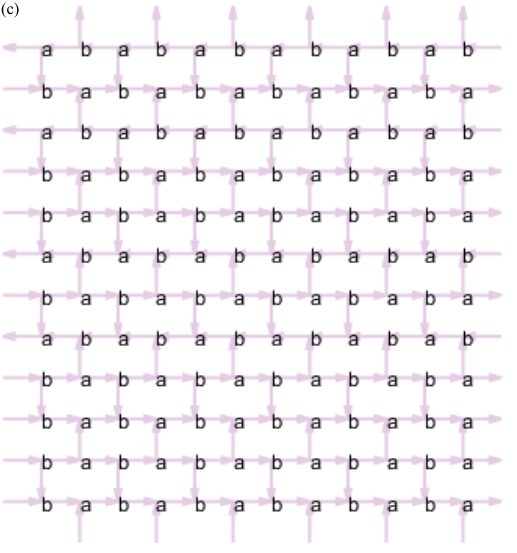}
    \includegraphics[width=0.48\textwidth,height=0.52\textwidth]{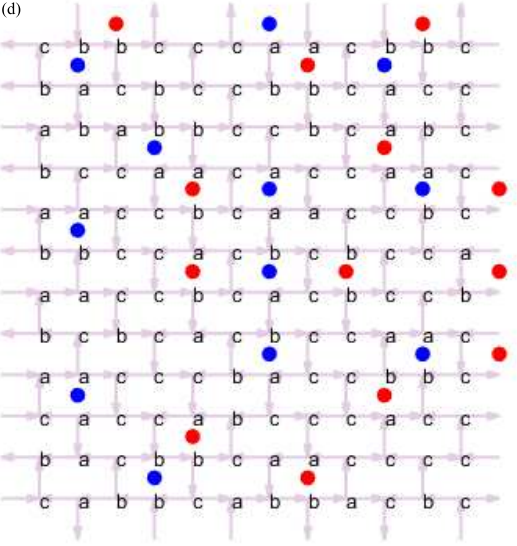}
    \includegraphics[width=0.98\textwidth,height=0.3\textwidth]{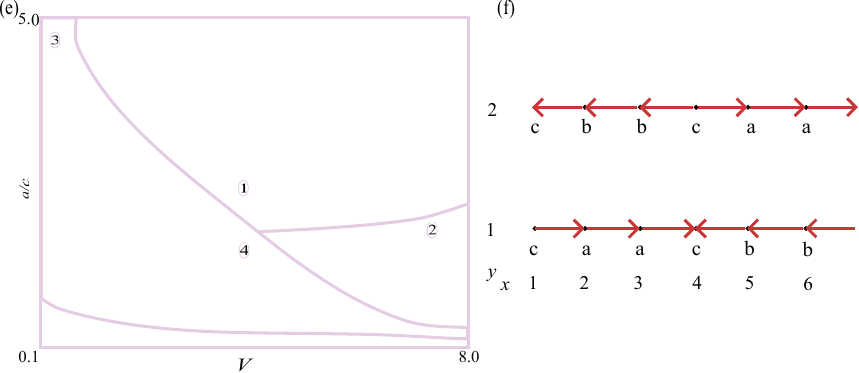}
    \caption{Snapshots for different phases. (a) $\rho_v=1/2$ striped vortex lattice (b) $\rho_v=2/3$ vortex lattice (c) \tcr{disordered} phase without vortex (d) \tcr{disordered} phase with vortex (e) the four snapshots are obtained at parameters marked in
        \ding{172}, \ding{173}, \ding{174} and \ding{175}. (f) The AFM pattern with \tcr{minimum} unit length 3 on the 2/3 VL lattice.}
    \label{fig:snapshot}
\end{figure*}



\bibliography{ref}
\end{document}